\documentclass[%
reprint,
 amsmath,amssymb,
 pre,
]{revtex4-1}

\usepackage{graphicx}
\usepackage{dcolumn}
\usepackage{bm}

\usepackage[all]{nowidow}
\usepackage{xcolor}
\DeclareMathAlphabet\mathbfcal{OMS}{cmsy}{b}{n}
\usepackage{mathrsfs} 
\newcommand{\parallelsum}{\mathbin{\!/\mkern-5mu/\!}}

\begin{document}


\title{Three-wave interactions in magnetized warm-fluid plasmas: \\ general theory with evaluable coupling coefficient}

\author{Yuan Shi}
\email{shi9@llnl.gov}
\affiliation{Lawrence Livermore National Laboratory, Livermore, CA 94550, USA}
 


\date{March 3, 2019}

\begin{abstract}
Resonant three-wave coupling is an important mechanism via which waves interact in a nonlinear medium. When the medium is a magnetized warm-fluid plasma, a previously-unknown formula for the coupling coefficients is derived by solving the fluid-Maxwell's equations to second order using multiscale perturbative expansions. The formula is not only general but also evaluable, whereby numerical values of the coupling coefficient can be determined for any three resonantly interacting waves propagating at arbitrary angles.
As one example, coupling coefficient governing laser scattering is evaluated. In conditions relevant to magnetized inertial confinement fusion experiments, lasers scatter from magnetized plasma waves and the growth rates are modified at oblique angles.
As another example, coupling coefficient between two Alfv\'en waves via a sound wave is evaluated. In conditions relevant to solar corona, the decay of a parallel Alfv\'en wave only slightly prefers exact backward geometry.

\end{abstract}

\maketitle


\section{Introduction\label{sec:intro}}
Plasmas are dielectric media wherein waves can interact nonlinearly. Unlike crystals whose optical properties may have limited range of tunability, plasma parameters can vary by many orders of magnitude. In particular, an adjustable optical axis can be introduced by applying an external magnetic field. Thereby, all nonlinear optical phenomena seen in crystals \cite{Bloembergen1996nonlinear} also occur in magnetized plasmas with ample flexibility. In addition to hosting optical phenomena, magnetized plasmas also support a zoo of other waves. These additional waves, such as the Alfv\'en wave, Bernstein waves, and hybrid waves, not only mediate new interactions between light waves, but also couples nonlinearly among themselves. For example, interactions between Alfv\'en waves is thought to be a major mechanism for anisotropic turbulence \cite{Sridhar1994toward,Ng1996interaction,Goldreich1997magnetohydrodynamic,Chandran2008weakly,Chaston2008turbulent,Gogoberidze2007nature,Schekochihin2012weak} and particle heating \cite{Voitenko05,Araneda2008proton} in astrophysical plasmas. 

While nonlinear wave coupling occurs in any dielectric medium, what makes the biggest difference is perhaps the coupling coefficient. When the coupling is weak, very large amplitude waves are needed in order to cause sizable effects. On the contrary, when the coupling is strong, even small amplitude waves can lead to observable consequences. Since plasma parameters span many orders of magnitude, it is impractical to exhaust the multidimensional parameter space by experiments and first-principle simulations. An analytical formula, which can be used to determine numerical values of the coupling coefficient, is thereof invaluable for mapping out wave-wave coupling behaviors in magnetized plasmas.  

For over half a century, numerous attempts are made to calculate wave coupling in magnetized plasmas due to three-wave interactions, which are the leading-order terms of the nonlinear response tensor \cite{Sagdeev1969nonlinear,Davidson72}. However, the presence of a background magnetic field significantly complicates the calculation, and most attempts start by restricting to a particular set of wave triad in some special geometry.
For example, theories have been developed when waves propagate perpendicular to the magnetic field \cite{Platzman1968light,Stenflo1972kinetic}, and explicit results have been obtained when the pump is the extraordinary wave \cite{Grebogi80,Grebogi1980parametric,Barr84,Boyd85,Simon1995parametric,Mourenas1996comment,Vyas16,Dodin2017parametric}, the ordinary wave \cite{Purohit2010excitation,Shi2017laser}, the upper-hybrid wave \cite{Ram82}, and the lower-hybrid wave \cite{Sanuki1977parametric}.
For wave propagation nearly parallel to the magnetic field, transverse and longitudinal modes decouple \cite{Sjolund1967non,Stenflo1970kinetic}, and results have been obtained when all waves are electrostatic \cite{Shivamoggi1982kinetic}, when the pump wave is a circularly polarized laser \cite{Laham1998effects}, the whistler wave \cite{Kumar2011stimulated}, the fast wave \cite{Voitenko2002nonlinear,Modi2013nonlinear}, and the Alfv\'en waves \cite{Hasegawa76,Hasegawa1976kinetic,Erokhin1978decay,Derby1978modulational,Goldstein1978instability,Wong1986parametric,Jayanti1993parametric,Matsukiyo2003parametric,Sweeney1978magneto,Brodin1988parametric,Brodin1990coupling,Vinas1991parallel,Jayanti1993dispersion,Hollweg1994beat,Voitenko1998three,Voitenko2000nonlinear,Shukla2004nonlinear,Ruderman2004stability,Brodin2006nonlinear,Nariyuki2007consequences}.
Although more general theories exist \cite{Galloway71,Boyd78,Liu1986parametric,Stenflo94,Brodin12}, the formal expressions of the coupling coefficient are too cumbersome to be useful and are rarely evaluated in practice \cite{Vinas1991oblique}. Moreover, in order to simplify results, increasing number of assumptions are usually made as the discussion progresses, and conflicting assumptions have led to numerous disputes in the literature.

In order to obtain a formula for the coupling coefficient that is not only general but also evaluable, a mathematically robust approach is necessary. In a previous paper \cite{Shi2017three}, an approached based on multiscale perturbative solution has been demonstrated for magnetized cold-fluid plasmas. The key to simplifying the general result is not to make additional assumptions, but to package seemingly complex terms into well-motivated operators. By studying properties of these mathematical operators, profound simplifications can then be unveiled, which would otherwise be buried under tedious arithmetics. This approach is not only useful for obtaining analytical expressions, but also necessary to avoid brute-force manipulation of large matrices during numerical evaluations.

In this paper, I will further demonstrate the operator approach by considering three-wave interactions in magnetized warm-fluid plasmas. The ideal warm-fluid model is applicable when the wavelengths of interest are much longer than the Debye length, while much shorter than the collisional mean free path. In this regime, plasma particles respond to perturbations collectively and dissipative effects are small. 
For fusion and astrophysical plasmas, the fluid-Maxwell model has a reasonable range of applicability. For example, in inertial confinement fusion conditions, the plasma density \mbox{$n\sim 10^{20}\,\text{cm}^{-3}$} and temperature \mbox{$T\sim 1$ keV}. Correspondingly, the Debye length $\lambda_D\sim 10^{-2}\,\mu\text{m}\, (T/n)^{1/2}$ is usually much smaller than the laser wavelength, which is in turn much smaller than the collisional mean free path $\lambda_{\text{mfp}}\sim 10\,\mu\text{m}\, (n Z^2)^{-1}$ in low-$Z$ plasmas.

Within the range of its applicability, the fluid model may then be solve perturbatively when amplitudes of fluctuations are small. To obtain solutions beyond the linear order, special procedures are necessary in order to avoid secular behaviors. A well-suited procedure is multiscale expansion, which expand spatial and temporal scales in addition to expanding the amplitudes. By renormalizing the spacetime,  well-behaved high-order perturbative solutions can then be obtained. In this weak-coupling regime, no further assumption is needed, and the model equations can be solved using rigorous procedures to study interactions between arbitrary waves in the most general geometry under a wide variety of plasma conditions.

This paper is organized as followes. In Sec.~\ref{sec:multi}, the fluid model and the multiscale method will be reviewed. In sec.~\ref{sec:linear}, I will introduce important operators and review linear waves from the operator perspective. In Sec.~\ref{sec:three}, I will derive the coupling coefficient by solving the second-order equations. In Sec.~\ref{sec:example}, known results in the literature will be recovered, and evaluation of the general formula will be demonstrated using two examples. Discussion is made in Sec.~\ref{sec:discussion} followed by a summary. Supplemental details are provided in the Appendix. 

\section{Warm-fluid model \label{sec:multi}}
The fluid model describes plasma species as charged gases, which couple with self-consistent electromagnetic fields through the Lorentz force law and the Maxwell's equations. The multi-fluids model can be regarded as moments of the kinetic model, and can be used to obtain magnetohydrodynhamics (MHD) models after further simplifications. 

\subsection{Fluid-Maxwell's equations\label{sec:equation}}
For each plasma species, its density evolves according to the continuity equation. In the absence of ionization and recombination, the number of particles is conserved, and the continuity equation is
\begin{equation}
    \label{eq:continuity}
    \frac{\partial n_s}{\partial t}+\nabla\cdot(n_s\mathbf{v}_s)=0,
\end{equation}
where $n_s$ is the density of species $s$, whose fluid velocity is $\mathbf{v}_s$. The continuity equation contains a nonlinear term $n_s\mathbf{v}_s$, which will contribute to wave-wave couplings. 

The fluid velocity evolves according to the momentum equation. Using the continuity equation, the nonrelativistic momentum equation can be written as
\begin{equation}
    \label{eq:momentum}
    m_s n_s\Big(\frac{\partial\mathbf{v}_s}{\partial t}+\mathbf{v}_s\cdot\nabla\mathbf{v}_s\Big)=-\nabla p_s+e_sn_s(\mathbf{E}+\mathbf{v}_s\times\mathbf{B}),
\end{equation}
where $m_s$ and $e_s$ are the mass and charge of each particle of species $s$, whose thermal motion leads to a pressure $p_s$. 
The above is the simplest momentum equation for warm plasmas, assuming collisions play negligible role, and the internal stress tensor $\tau_{ij}=-p\delta_{ij}$ remains isotropic despite of external forces. 

To close the infinite hierarchy of fluid equations, we can express the pressure in terms of density and velocity. For simplicity, consider polytropic process for which $p V^\xi$ is a constant, where $\xi\ge 0$ is the polytropic index, $p$ is the pressure, and $V$ is the volume of the fluid element. 
Suppose the number of particles in the fluid element is constant, then the polytropic condition relates changes of pressure and density by
\begin{equation}
    \label{eq:pressure}
    n_s d_t p_s=\xi_s p_s d_tn_s,
\end{equation}
where $d_t=\partial_t+\mathbf{v}_s\cdot\nabla$ is the convective derivative. 
The polytropic process assumes that the heat to work ratio is a constant. In particular, the polytropic process recovers the isobaric process when $\xi=0$; the isothermal process when $\xi=1$; the isochoric process when $\xi=\infty$; and the adiabatic process when $\xi=C_p/C_v$, where $C_p$ and $C_v$ are heat capacities at constant pressure and volume.

To model plasmas with self-consistent electric and magnetic fields, we can couple the fluid equations with the Maxwell's equations. The time evolution of the magnetic field is given by the Faraday's law:
\begin{equation}
    \label{eq:Faraday}
    \frac{\partial \mathbf{B}}{\partial t}=-\nabla\times\mathbf{E},
\end{equation}
which is independent of plasma dynamics. In comparison, the time evolution of the electric field is given by the Amp\`ere's law:
\begin{equation}
    \label{eq:ampere}
    \frac{\partial \mathbf{E}}{\partial t}=c^2\nabla\times\mathbf{B}-\frac{1}{\epsilon_0}\sum_s e_sn_s\mathbf{v}_s,
\end{equation}
where $\epsilon_0$ is the vacuum permittivity. The other two Maxwell's equations $\nabla\cdot\mathbf{E}=\sum_s e_s n_s/\epsilon_0$ and $\nabla\cdot\mathbf{B}=0$ are guaranteed once they are satisfied at the initial time. 

The fluid-Maxwell system satisfies local energy-momentum conservation laws. The total energy density of the system is 
\begin{equation}
    \label{eq:energy}
    U=\frac{1}{2}\epsilon_0\mathbf{E}^2+\frac{1}{2\mu_0}\mathbf{B}^2+ \sum_s\frac{1}{2}\Big(m_sn_s\mathbf{v}_s^2+\frac{p_s}{\xi_s-1}\Big),
\end{equation}
which is the sum of the field, the kinetic, and the thermal energy densities. Similarly, the energy flux is 
\begin{equation}
    \label{eq:energyflux}
    \mathbf{S}=\frac{1}{\mu_0}\mathbf{E}\times\mathbf{B} +\sum_s\frac{1}{2}\Big(m_sn_s\mathbf{v}_s^2\mathbf{v}_s +\frac{\xi_s}{\xi_s-1}p_s\mathbf{v}_s\Big),
\end{equation}
which is constituted of the Poynting flux, the kinetic flux, and the thermal flux. The local conservation law is
\begin{equation}
    \label{eq:conserve}
    \partial_t U+\nabla\cdot\mathbf{S}=0,
\end{equation}
which can be verified by straightforward calculations using Eqs.~(\ref{eq:continuity})-(\ref{eq:ampere}). Analogously, one can show that the local momentum is also conserved: $\partial_t \Pi_j+\partial_{i} \sigma_{ij}=0$, where $\Pi_i$ is the momentum density, and $\sigma_{ij}$ is the stress tensor.

\subsection{Multiscale perturbative expansions\label{sec:solution}}
The fluid-Maxwell's equations are a set of nonlinear partial differential equations. The equations self-consistently determine the fluid variables $n_s$, $\mathbf{v}_s$, and $p_s$, as well as the field variables $\mathbf{E}$ and $\mathbf{B}$. Although calculating the general solution is difficult, perturbative solutions may be obtained when fluctuations have small amplitudes. 
Since the equations are nonlinear, special procedures are needed in order to remove secular behaviors beyond the leading order. Once secular behaviors are removed, the perturbative solutions are well-behaved without violating the small-amplitude assumption.

Consider perturbations from an equilibrium state of the plasma. Then, a generic fluid or field variable $\mathbf{Z}$ can be expanded as
\begin{eqnarray}
    \label{eq:expand}
    \mathbf{Z}&=&\mathbf{Z}_0+\lambda\mathbf{Z}_1+\lambda^2\mathbf{Z}_2+\dots.
\end{eqnarray}
Here, the equilibrium state is labeled by the subscript ``0", and $\lambda$ is an auxiliary smallness parameter. 
Notice that at this step, it is not necessary to assumed any property of higher order terms $\mathbf{Z}_j$. In particular, the average $\langle \mathbf{Z}_j \rangle$ needs not be zero, and the equilibrium state $\mathbf{Z}_0$ is not necessarily the averaged quantity.

One way of removing secular behavior from the perturbative solution is to also expand temporal and spatial scales. Using multiscale expansions, the time and space derivatives are
\begin{eqnarray}
\label{eq:tscale}
    \partial_t&=&\partial_{t_0}+\lambda\partial_{t_1}+\lambda^2\partial_{t_2}+\dots,\\
    \label{eq:xscale}
    \partial_i&=&\partial_{i_0}+\lambda\partial_{i_1}+\lambda^2\partial_{i_2}+\dots,
\end{eqnarray}
where $\lambda$ is the same auxiliary expansion parameter. The multiscale expansion assumes that weaker interactions occur on slower time scales and larger spatial scales. This is intuitive because weaker couplings require further accumulations before their effects become appreciable. 
In the multiscale expansion, $t_1$ is a time scale slower than $t_0$ by a factor of $\lambda$, and processes that occur on $t_1$ scale is assumed to be well separated from processes on $t_0$ scale. Similarly, other temporal and spatial scales are ordered by $\lambda$ and are assumed to be independent.

Consider the simplest equilibrium where the plasma is uniform and stationary under a constant background magnetic field. In this case, $\mathbf{E}_0$ and $\mathbf{v}_{s0}$ are zero, whereas $\mathbf{B}_0$, $n_{s0}$, and $p_{s0}$ are nonzero but constant. Although this simple situation is rarely encountered in realistic plasmas, it provides a reasonable simplification when the scales of inhomogeneities are well-separated from characteristic scales of three-wave interactions.

When perturbed from the simple equilibrium, the $\lambda$-order equations are homogeneous linear partial differential equations with constant coefficients. 
The linearized fluid equations are
\begin{eqnarray}
    \label{eq:n1}
    \partial_{t_0} n_{s1}&=&-n_{s0}\nabla_0\cdot\mathbf{v}_{s1},\\
     \label{eq:v1}
    m_sn_{s0}\partial_{t_0}\mathbf{v}_{s1}&=&-\nabla_0 p_{s1}+e_sn_{s0}(\mathbf{E}_1+\mathbf{v}_{s1}\times\mathbf{B}_0),\quad \\
    \label{eq:p1}
    n_{s0}\partial_{t_0}p_{s1}&=&\xi_sp_{s0}\partial_{t_0}n_{s1},
\end{eqnarray}
which couple $n_{s1}$, $\mathbf{v}_{s1}$, and $p_{s1}$ in pairs. 
The linearized Maxwell's equations are
\begin{eqnarray}
    \label{eq:B1}
    \partial_{t_0}\mathbf{B}_1&=&-\nabla_0\times\mathbf{E}_1,\\
    \label{eq:E1}
    \partial_{t_0}\mathbf{E}_1&=&c^2\nabla_0\times\mathbf{B}_1-\frac{1}{\epsilon_0}\sum_s e_sn_{s0}\mathbf{v}_{s1}.
\end{eqnarray}
These linear partial differential equations can be easily solves in the Fourier space to obtain the full spectrum of linear waves in magnetized warm-fluid plasmas.

To compute three-wave coupling between linear waves, we need to solve the equation to the next order. The $\lambda^2$-order continuity equation is
\begin{eqnarray}
    \label{eq:n2}
    \nonumber
    &&\partial_{t_0}n_{s2}+n_{s0}\nabla_0\cdot\mathbf{v}_{s2}\\
    &=&-\partial_{t_1}n_{s1}-n_{s0}\nabla_1\cdot\mathbf{v}_{s1}-\nabla_0\cdot(n_{s1}\mathbf{v}_{s1}),
\end{eqnarray}
the $\lambda^2$-order momentum equation is
\begin{eqnarray}
    \label{eq:v2}
    \nonumber
    &&m_sn_{s0}\partial_{t_0}\mathbf{v}_{s2}+\nabla_0p_{s2}-e_sn_{s0}(\mathbf{E}_2+\mathbf{v}_{s2}\times\mathbf{B}_0)\\
     \nonumber
    &=&-m_s[n_{s0}(\partial_{t_1}\!\mathbf{v}_{s1}\!+\!\mathbf{v}_{s1}\!\cdot\!\nabla_0\mathbf{v}_{s1}\!)\!+\!n_{s1}\partial_{t_0}\mathbf{v}_{s1}]\!-\!\nabla_1 p_{s1}\hspace{3pt}\\
    &&+e_s[n_{s0}\mathbf{v}_{s1}\times\mathbf{B}_1+n_{s1}(\mathbf{E}_1+\mathbf{v}_{s1}\times\mathbf{B}_0)],
\end{eqnarray}
and the $\lambda^2$-order pressure equation is
\begin{eqnarray}
    \label{eq:p2}
    \nonumber
    &&n_{s0}\partial_{t_0}p_{s2}-\xi_s p_{s0}\partial_{t_0}n_{s2}\\
    \nonumber
    &=&-n_{s0}(\partial_{t_1}p_{s1}+\mathbf{v}_{s1}\cdot\nabla_0 p_{s1})-n_{s1}\partial_{t_0}p_{s1}\\
    &&+\xi_s [p_{s0}(\partial_{t_1}n_{s1}+\mathbf{v}_{s1}\cdot\nabla_0n_{s1})+ p_{s1}\partial_{t_0}n_{s1}].
\end{eqnarray}
The above equations may be simplified using $\lambda$-order equations, as will be done later in Sec.~\ref{sec:second}.
Similarly, we can write down the $\lambda^2$-order Maxwell's equation. The second-order Faraday's law is
\begin{eqnarray}
    \label{eq:B2}
    \partial_{t_0}\mathbf{B}_2+\nabla_0\times\mathbf{E}_2=-\partial_{t_1}\mathbf{B}_1-\nabla_1\times\mathbf{E}_1,
\end{eqnarray}
and the second-order Amp\`ere's law is
\begin{eqnarray}
    \label{eq:E2}
    \nonumber
    &&\partial_{t_0}\mathbf{E}_2-c^2\nabla_0\times\mathbf{B}_2+\frac{1}{\epsilon_0}\sum_s e_sn_{s0}\mathbf{v}_{s2}\\
    &=&-\partial_{t_1}\mathbf{E}_1+c^2\nabla_1\times\mathbf{B}_1-\frac{1}{\epsilon_0}\sum_s e_sn_{s1}\mathbf{v}_{s1}.
\end{eqnarray}
Although these equations may look complicated, they are in fact linear equations for fluid variables $n_{s2}$, $\mathbf{v}_{s2}$, and $p_{s2}$, as well as field variables $\mathbf{E}_2$ and $\mathbf{B}_2$.
Moreover, these second order variables couple in exactly the same way as in the $\lambda$-order equations. The only difference is the presence of source terms, which I have arranged to appear on the right-hand-sides (RHS) of the above equations. Once the first-order variables are solved from the $\lambda$-order equations, these source terms can be regarded as known. The above $\lambda^2$-order equations are then a system of inhomogeneous linear partial differential equations, which can be solved again in the Fourier space.

\section{Magnetized linear waves\label{sec:linear}}
Before discussing three-wave interactions, it is useful to familiarize with linear waves in magnetized warm-fluid plasmas. In this section, I will review the eigenvalues, the eigenvectors, and the eigenenergies of linear waves. During this review, I will also introduce important concepts that will become indispensable in the next section.

\subsection{First-order equations\label{sec:first}}
Now let us solve the first-order equations. Since the equations are linear, the general solution is a superposition of plane waves. In particular, the first-order electric and magnetic fields are given by
\begin{eqnarray}
    \label{eq:E1wave}
    \mathbf{E}_1&=&\frac{1}{2}\sum_{\mathbf{k}\in\mathbb{K}_1}\mathbfcal{E}_{1,\mathbf{k}} e^{i\theta_{\mathbf{k}}},\\
    \label{eq:B1wave}
    \mathbf{B}_1&=&\frac{1}{2}\sum_{\mathbf{k}\in\mathbb{K}_1}\frac{\mathbf{k}\times\mathbfcal{E}_{1,\mathbf{k}}}{\omega_\mathbf{k}} e^{i\theta_{\mathbf{k}}},
\end{eqnarray}
where Faraday's law has been used to related $\mathbf{B}_1$ to $\mathbf{E}_1$. In the above spectral expansion, 
$\theta_{\mathbf{k}}=\mathbf{k}\cdot\mathbf{x}_0-\omega_\mathbf{k}t_0$ is the fast varying phase, $\mathbf{\mathbfcal{E}}_{1,\mathbf{k}}(\mathbf{x}_1,t_1,\dots)$ is the slowly varying amplitude, and the summation is over a discrete spectrum $\mathbb{K}_1$. Since the electric field is real-valued, whenever $\mathbf{k}\in\mathbb{K}_1$, we must also have $-\mathbf{k}\in\mathbb{K}_1$. Moreover, we need the reality conditions $\omega_{-\mathbf{k}}=-\omega_{\mathbf{k}}$ and $\mathbf{\mathbfcal{E}}_{-\mathbf{k}}=\mathbf{\mathbfcal{E}}_{\mathbf{k}}^*$, where the star denotes complex conjugation. It is easy to check that once these conditions are satisfied, $\mathbf{B}_1$ is also real-valued.

The three fluid variables can also be expressed in terms of the electric field. The pressure equation is easy to solve, which gives a simple linear relation
\begin{equation}
    \label{eq:p1wave}
    p_{s1}=\varepsilon_s n_{s1},
\end{equation}
where the constant $\varepsilon_s:=\xi_s p_{s0}/n_{s0}$ has the unit of energy. Assuming ideal gas law $p_0=n_0k_B T_0$, then $\varepsilon=\xi k_B T_0$ is proportional to the temperature.
Substituting the above relation into the momentum equation, $\mathbf{v}_{s1}$ and $n_{s1}$ can be solved in conjunction with the continuity equation:
\begin{eqnarray}
    \label{eq:v1wave}
    \mathbf{v}_{s1}&=&\frac{ie_s}{2m_s} \sum_{\mathbf{k}\in\mathbb{K}_1} \frac{\hat{\mathbb{F}}_{s,\mathbf{k}}\mathbfcal{E}_{1,\mathbf{k}}}{\omega_\mathbf{k}} e^{i\theta_{\mathbf{k}}},\\
    \label{eq:n1wave}
    n_{s1}&=&\frac{ie_sn_{s0}}{2m_s} \sum_{\mathbf{k}\in\mathbb{K}_1} \frac{\mathbf{k}\cdot\hat{\mathbb{F}}_{s,\mathbf{k}}\mathbfcal{E}_{1,\mathbf{k}}}{\omega^2_\mathbf{k}} e^{i\theta_{\mathbf{k}}}.
\end{eqnarray}
Here, the solution is expressed in terms the warm forcing operator $\hat{\mathbb{F}}_{s,\mathbf{k}}:\mathbb{C}^3\rightarrow\mathbb{C}^3$, which is a linear operator and is specific to each species and wave vector.

To convert the above symbolic expressions to actual solutions, we need to find an explicit expression for the warm forcing operator.
Using the momentum equation, the warm forcing operator satisfies 
\begin{equation}
    \label{eq:FhatZ}
    \hat{\mathbb{F}}_{s,\mathbf{k}}\mathbf{Z}=\mathbf{Z}+i\beta_{s,\mathbf{k}} (\hat{\mathbb{F}}_{s,\mathbf{k}}\mathbf{Z})\times\mathbf{b}+\frac{u_s^2}{\omega^2_\mathbf{k}}\mathbf{k}(\mathbf{k}\cdot\hat{\mathbb{F}}_{s,\mathbf{k}}\mathbf{Z}),
\end{equation}
for any $\mathbf{Z}\in\mathbb{C}^3$. In the above equation, $\beta_{s,\mathbf{k}}=\Omega_s/\omega_\mathbf{k}$ is the magnetization ratio, where $\Omega_s=e_sB_0/m_s$ is the gyrofrequency; $\mathbf{b}$ is the unit vector along $\mathbf{B}_0$; and $u_s^2:=\varepsilon_s/m_s=\xi_s k_B T_{s0}/m_s$ is the thermal speed.
It is easy to see that the inverse operator satisfies $\hat{\mathbb{F}}^{-1}_{ij}=\delta_{ij}-i\beta\epsilon_{ijl}b_l-u^2k_ik_j/\omega^2$, where $\epsilon_{ijl}$ is the Levi-Civita symbol and I have abbreviated all subscripts for simplicity. 
Inverting $\hat{\mathbb{F}}^{-1}$, the forcing operator can be expressed as the composite:
\begin{equation}
    \label{eq:Fhat}
    \hat{\mathbb{F}}=\mathbb{F} \mathbb{P}=\mathbb{P}^\dagger \mathbb{F},
\end{equation}
where $\mathbb{F}$ is the cold forcing operator and $\mathbb{P}$ is the pressure operator.
The cold forcing operator acts on any complex vector by \cite{Shi2017three}
\begin{equation}
    \label{eq:F}
    \mathbb{F}\mathbf{Z}=\gamma^2[\mathbf{Z}+i\beta\mathbf{Z}\times\mathbf{b}-\beta^2(\mathbf{Z}\cdot\mathbf{b})\mathbf{b}],
\end{equation}
where $\gamma^2=1/(1-\beta^2)$ is the magnetization factor. It is easy to check that $\mathbb{F}$ recovers the identity operator in the unmagnetized limit.
The pressure operator acts on any complex vector by
\begin{equation}
    \label{eq:Popt}
    \mathbb{P}\mathbf{Z}=\mathbf{Z}+\hat{\gamma}^2\frac{u^2}{\omega^2}\mathbf{k}(\mathbf{k}\cdot\mathbb{F}\mathbf{Z}),
\end{equation}
where $\hat{\gamma}^2=1/(1-\hat{\beta}^2)$ is the thermal factor and $\hat{\beta}^2=u^2(\mathbf{k}\cdot\mathbb{F}\mathbf{k})/\omega^2$ is the thermal ratio. It is easy to check that $\mathbb{P}$ recovers the identity operator in the cold limit.
Using $\mathbf{k}\cdot\mathbb{F}\mathbf{k}=\gamma^2[\mathbf{k}^2-\beta^2(\mathbf{k}\cdot\mathbf{b})^2]$, it is a straightforward calculation to verify that $\hat{\mathbb{F}}$ given by the above formulas satisfies Eq.~(\ref{eq:FhatZ}).
The warm forcing operator inherits a number of properties from $\mathbb{F}$ and $\mathbb{P}$. First, since $\mathbb{F}^\dagger=\mathbb{F}$ is self-adjoint with respect to vector inner products, the warm forcing operator $\hat{\mathbb{F}}^\dagger=\hat{\mathbb{F}}$ is also self-adjoint, although \mbox{$\mathbb{P}^\dagger\ne\mathbb{P}$} is not.
Second, since $\mathbb{F}_{-\mathbf{k}}=\mathbb{F}^*_{\mathbf{k}}$, the pressure operator $\mathbb{P}_{-\mathbf{k}}=\mathbb{P}^*_{\mathbf{k}}$ and the warm forcing operator $\hat{\mathbb{F}}_{-\mathbf{k}}=\hat{\mathbb{F}}^*_{\mathbf{k}}$. It is then easy to see that the fluid variables given by Eqs.(\ref{eq:p1wave})-(\ref{eq:n1wave}) are real-valued.

Having expressed all fluctuations in terms of $\mathbfcal{E}_{1,\mathbf{k}}$, the only remaining equation is the Amp\`ere's law, which can be use to constrain the electric field. Substituting Eqs.~(\ref{eq:B1wave}) and (\ref{eq:v1wave}) into Eq.~(\ref{eq:E1}), each Fourier component satisfies the matrix equation $\mathbb{D}_\mathbf{k} \mathbfcal{E}_{1,\mathbf{k}}=\mathbf{0}$,
where the dispersion tensor
\begin{equation}
    \label{eq:Dtensor}
    \mathbb{D}_\mathbf{k}^{ij}=(\omega_\mathbf{k}^2-c^2\mathbf{k}^2)\delta^{ij}+c^2k^i k^j-\sum_s\omega_{ps}^2\hat{\mathbb{F}}_{s,\mathbf{k}}^{ij}.
\end{equation}
Here, $\omega_{ps}^2=e_s^2n_{s0}/\epsilon_0m_s$ is the plasma frequency of species $s$. From the above first-order electric-field equation, it is easy to see that the forcing operator $\hat{\mathbb{F}}_{s,\mathbf{k}}$ is related to the linear susceptibility by
\begin{equation}
    \label{eq:chi}
    \chi_{s,\mathbf{k}}=-\frac{\omega_{ps}^2}{\omega_\mathbf{k}^2}\hat{\mathbb{F}}_{s,\mathbf{k}}.
\end{equation}
Although the susceptibility is commonly used in linear wave theories, the forcing operator is more convenient when discussing nonlinear wave-wave couplings.

\subsection{Dispersion relations\label{sec:dispersion}}
The first-order electric-field equation has nonzero solutions if and only if the dispersion tensor is degenerate. The degeneracy condition gives the dispersion relation $\det\mathbb{D}_\mathbf{k}=\mathbf{0}$,
which constrains the wave frequency $\omega_\mathbf{k}$ as a function of the wavevector. For each wavevector $\mathbf{k}$, there can be multiple solutions of $\omega_\mathbf{k}$, each living on a separate dispersion branch.

When evaluating determinant of the dispersion tensor, it is convenient to use its matrix representations. A particularly convenient coordinate is the field coordinate $(\hat{\mathbf{x}},\hat{\mathbf{y}},\hat{\mathbf{z}})$, in which $\mathbf{B}_0=(0,0,B_0)$ is aligned with $\hat{\mathbf{z}}$ and $\mathbf{k}=k(\sin\theta,0,\cos\theta)$ is in the $x$-$z$ plane. In this coordinate, the warm forcing operator $\hat{\mathbb{F}}$ is represented by the Hermitian matrix
\begin{eqnarray}
    \nonumber
    \hat{\mathcal{F}}\!=\!
    \left(\hspace{-5pt} \begin{array}{ccc}
    \gamma^2(1\!+\!\gamma^2\rho^2\mathrm{s}^2_\theta) \!&\! i\beta\gamma^2(1\!+\!\gamma^2\rho^2\mathrm{s}^2_\theta) \!&\! \gamma^2\rho^2\mathrm{s}_\theta\mathrm{c}_\theta \\
    -i\beta\gamma^2(1\!+\!\gamma^2\rho^2\mathrm{s}^2_\theta)  \!&\! \gamma^2(1\!+\!\beta^2\gamma^2\rho^2\mathrm{s}^2_\theta) \!&\! -i\beta\gamma^2\rho^2\mathrm{s}_\theta\mathrm{c}_\theta \\
    \gamma^2\rho^2\mathrm{s}_\theta\mathrm{c}_\theta \!&\! i\beta\gamma^2\rho^2\mathrm{s}_\theta\mathrm{c}_\theta \!&\! 1\!+\!\rho^2\mathrm{c}^2_\theta
    \end{array}\hspace{-5pt} \right)\!,
\end{eqnarray}
where the reduced thermal factor $\rho^2=\hat{\gamma}^2u^2k^2/\omega^2$, and I have abbreviated $\mathrm{s}_\theta:=\sin\theta$ and $\mathrm{c}_\theta:=\cos\theta$. 
Summing over responses of all plasma species, the dispersion tensor is represented by the matrix
\begin{eqnarray}
    \nonumber
    \frac{\mathcal{D}}{\omega^2}\! =\!
    \left(\hspace{-3pt} \begin{array}{ccc}
    S\!-\!n^2\mathrm{c}^2_\theta & -iD & (n^2-T)\mathrm{s}_\theta\mathrm{c}_\theta \\
    iD  & S\!+\!T\mathrm{s}^2_\theta\!-\!n^2 & iE\mathrm{s}_\theta\mathrm{c}_\theta \\
    (n^2\!-\!T)\mathrm{s}_\theta\mathrm{c}_\theta & -iE\mathrm{s}_\theta\mathrm{c}_\theta & P\!-\!n^2\mathrm{s}^2_\theta
    \end{array}\hspace{-3pt} \right),\hspace{6pt}
\end{eqnarray}
where $n=ck/\omega$ is the refractive index, and components of the dielectric tensor are related to
\begin{eqnarray}
    \label{eq:S}
    S&=&1-\sum_s\frac{\omega_{ps}^2}{\omega^2}\gamma^2_s(1+\gamma_s^2\rho_s^2\mathrm{s}^2_\theta ),\\
    \label{eq:D}
    D&=&\sum_s\frac{\omega_{ps}^2}{\omega^2}\beta_s\gamma^2_s(1+\gamma_s^2\rho_s^2\mathrm{s}^2_\theta ),\\
    \label{eq:P}
    P&=&1-\sum_s\frac{\omega_{ps}^2}{\omega^2}(1+\rho_s^2\mathrm{c}^2_\theta ),\\
    \label{eq:T}
    T&=&\sum_s\frac{\omega_{ps}^2}{\omega^2}\gamma^2_s\rho_s^2,\\
    \label{eq:E}
    E&=&\sum_s\frac{\omega_{ps}^2}{\omega^2}\beta_s\gamma^2_s\rho_s^2.
\end{eqnarray}
The above expressions recover the standard Stix symbols in the cold limit, where $\rho^2$ becomes zero and the $k$-dependence of the dielectric tensor vanishes.

Taking determinant of the dispersion matrix, the wave dispersion relation can be written in the form
\begin{equation}
    \label{eq:disp}
    A n^4-B n^2+C=0.
\end{equation}
Coefficients in the above equation depend on $n^2$ as well as $\omega$ due to thermal effects:
\begin{eqnarray}
    \label{eq:A}
    A&=&S' \mathrm{s}^2_\theta+P'\mathrm{c}^2_\theta,\\
    \label{eq:B}
    B&=&R'L'\mathrm{s}^2_\theta+S'P'(1+\mathrm{c}^2_\theta)+2TA,\\
    \label{eq:C}
    C&=&P'R'L'+T(B-TA)+E(2P'D'-EA)\mathrm{c}^2_\theta.\hspace{5pt}
\end{eqnarray}
Here, $S'=S-T\mathrm{c}^2_\theta$, $ D'=D+E\mathrm{c}^2_\theta$, and $P'=P-T \mathrm{s}^2_\theta$. Analogous to the cold case, $R'=S'+D'$ and $L'=S'-D'$. If we formally treat Eq.~(\ref{eq:disp}) as a quadratic equation for $n^2$, the determinant $F^2=B^2-4AC=(R'L'-S'P')^2\mathrm{s}^4_\theta+4(D'P'-EA)^2\mathrm{c}^2_\theta\ge 0$. The formal solutions $n^2=(B\pm F)/2A$ then give two implicit equations for $n^2$ as a function of the wave frequency.
In the limit $ck\rightarrow 0$, thermal effects vanish, and the asymptotic dispersion relation is discussed in Appendix~\ref{app:low}.

\begin{figure}[t]
    \centering
    \includegraphics[width=0.48\textwidth]{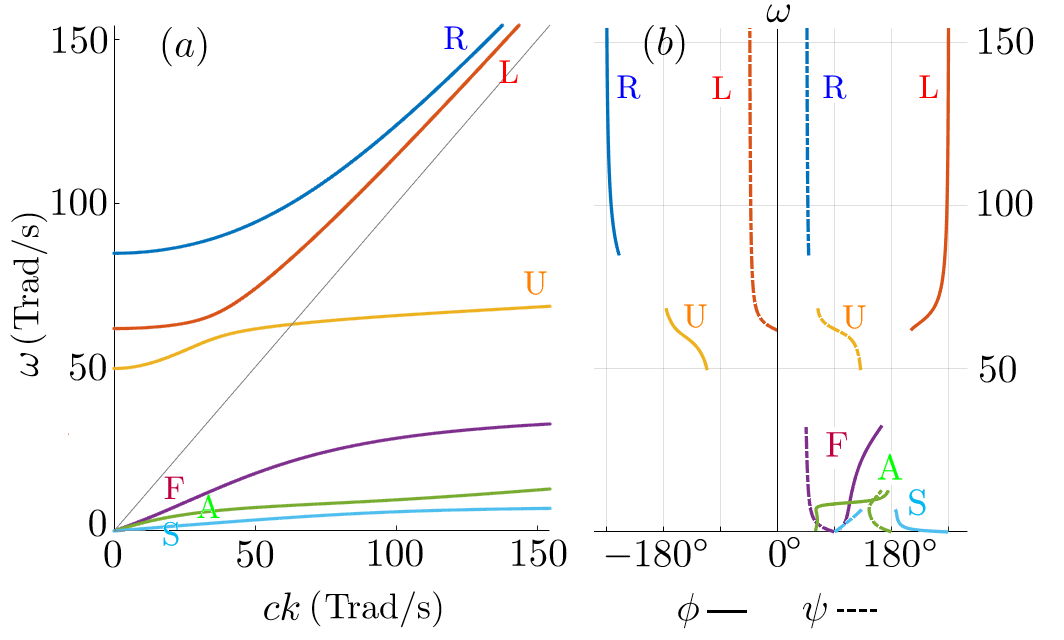}
    \caption{Wave dispersion relations (a) and polarization angles (b) in magnetized warm-fluid electron-ion plasma when \mbox{$\langle \mathbf{k},\mathbf{B}_0\rangle=30^\circ$}.
    The two electromagnetic (EM) waves are elliptically polarized R wave (blue) and L wave (red), which become transverse and approach the light cone $\omega=ck$ when $ck\rightarrow\infty$. 
    The other four branches are plasma waves, which become longitudinal when $ck\rightarrow\infty$. In this limit and when $\langle \mathbf{k},\mathbf{B}_0\rangle\rightarrow 90^\circ$, the yellow branch is the upper-hybrid (UH) wave and the purple branch is the lower-hybrid (LH) wave.
    In the opposite limit $ck\rightarrow 0$, the purple branch is the fast (F) wave, the green branch is the Alfv\'en (A) wave, and the cyan branch is the slow (S) wave. 
    For all dispersion branches to be visible on the same scale ($10^{12}$~rad/s), the mass ratio $m_i/m_e=5$ is artificial. The plasma density is $n_e=n_i=10^{18} \;\text{cm}^{-3}$; the plasma temperature is $T_e=T_i=3.2$ keV; the polytropic index is adiabatic $\xi_e=\xi_i=3$; the magnetic field is \mbox{$B_0=2.5$ MG} such that $|\Omega_e|/\omega_{pe}\approx 0.8$ and $v_A/c_s\approx 4$, where $v_A$ is the Alfv\'en speed and $c_s$ is the sound speed. 
    }
    \label{fig:wave}
\end{figure}

A numerically robust procedure for evaluating the dispersion relation is to converted it to a polynomial equation for $\omega^2$, using which wave frequencies can be solved as functions of the wavevector \mbox{(Fig.~\ref{fig:wave}a)}. 
To see what multiplicative prefactor is needed, notice that the rational functions $A$, $B$, and $C$ have a pole at \mbox{$\omega^2=0$}. In addition, each warm species contribute two poles at \mbox{$\omega^4-(\Omega_s^2+u_s^2k^2)\omega^2+\Omega_s^2u_s^2k^2\mathrm{c}^2_\theta=0$}. One of these poles becomes degenerate with the $\omega^2=0$ pole either when the species is cold, in which case the other pole becomes the magnetic pole \mbox{$\omega^2-\Omega_s^2=0$}, or when the species is unmagnetized, in which case the other pole becomes the thermal pole \mbox{$\omega^2-u_s^2 k^2=0$}. 
For parallel wave propagation, $\mathrm{c}^2_\theta=1$, so the magnetic and the thermal poles decouple; for perpendicular wave propagation, $\mathrm{c}^2_\theta=0$, so one pole becomes $\omega^2=0$ while the other pole becomes the hybrid pole \mbox{$\omega^2-\Omega_s^2-u_s^2 k^2=0$}; the two poles are otherwise mixed at general angles of propagation.
After multiplying the minimal pole-removing prefactor on both sides of Eq.~(\ref{eq:disp}), it becomes a polynomial equation for $\omega^2$ of degree $N$. 
For an unmagnetized plasma $N=3+N_c$, where $N_c=N_t+\text{sgn}(N_s-N_t)-1$ is the number of sound waves. Here, $N_s$ is the total number of plasma species, $N_t$ is the number of warm species, and $\text{sgn}$ is the sign function.
When the plasma becomes magnetized, $N=3+N_c+N_s$, because each species contributes an additional cyclotron resonance.
At general propagation angles, the dispersion relation is constituted of $N$ separate branches with hybrid characteristics.

\subsection{Polarization of eigenmodes \label{sec:polarization}}
Once the dispersion relation is satisfied, the first-order electric-field equation has nontrivial solutions. The solution space is a one-dimensional vector space when $\omega_\mathbf{k}$ is nondegenerate. In this case, the vector space is $\mathbfcal{E}_{1,\mathbf{k}}=\mathcal{E}_{1,\mathbf{k}}\mathbf{e}_\mathbf{k}$, where $\mathcal{E}_{1,\mathbf{k}}\in\mathbb{C}$ is an arbitrary complex scalar and the unit polarization vector $\mathbf{e}^\dagger_\mathbf{k}\mathbf{e}_\mathbf{k}=1$ is completely specified, up to the $U(1)$ symmetry, by two polarization angles on the unit sphere. 

It is physically meaningful to specify the two polarization angles in relation to the wavevector and the magnetic field. When $\mathbf{k}$ and $\mathbf{B}_0$ are not aligned, the unit vector can be decomposed as $\mathbf{e}=\hat{\mathbf{k}}e_k-i\hat{\mathbf{y}}e_y+(\hat{\mathbf{k}}\times\hat{\mathbf{y}})e_\times$, where the unit vector 
$\hat{\mathbf{y}}\parallelsum\mathbf{B}_0\times\mathbf{k}$. In spherical coordinate, components of $\mathbf{e}$ can be written as $e_k=\cos\phi$, $e_y=\sin\phi\sin\psi$, and $e_\times=\sin\phi\cos\psi$.
The wave is longitudinal when $\phi=0^\circ$ and transverse when $\phi=90^\circ$; the wave electric field is in the $\mathbf{k}$-$\mathbf{B}_0$ plane when $\psi=0^\circ$ and at maximum angle with the plane when $\psi=90^\circ$.
Since arbitrary scaling is allowed, the unit vector $\mathbf{e}$ is defined on the projective space. In terms of $\phi$ and $\psi$, the wave polarization is invariant under transformations $\Psi_{\pm}:(\phi,\psi)\rightarrow(-\phi,\psi\pm 180^\circ)$ and $\Phi_{\pm}:(\phi,\psi)\rightarrow(\phi\pm 180^\circ,\psi)$. since the polarization angles are periodic in $360^\circ$, the wave polarization is invariant under actions of $\Psi_{\pm}$ and $\Phi_{\pm}$ in arbitrary compositions.

To compute polarization angles for each eigenmode, it is more convenient to use the wave coordinate, 
which is related to the field coordinate by $(\hat{\mathbf{k}},-i\hat{\mathbf{y}},\hat{\mathbf{k}}\times\hat{\mathbf{y}})=(\hat{\mathbf{x}},\hat{\mathbf{y}},\hat{\mathbf{z}})\mathcal{L}_y(\theta)$, where $\mathcal{L}_y(\theta)$ is a left-handed rotation around $\hat{\mathbf{y}}\rightarrow-i\hat{\mathbf{y}}$ by angle $\theta$. 
In the wave coordinate, the dispersion tensor is represented by a different matrix $\mathcal{D}'=\mathcal{L}^{-1}\mathcal{D}\mathcal{L}$, which can be written explicitly as
\begin{eqnarray}
    \nonumber
    \frac{\mathcal{D}'}{\omega^2}\! =\!
    \left(\hspace{-4pt} \begin{array}{ccc}
    S'\mathrm{s}^2_\theta\!+\!P'\mathrm{c}^2_\theta & -D'\mathrm{s}_\theta & (P'\!-\!S')\mathrm{s}_\theta\mathrm{c}_\theta \\
    -D'\mathrm{s}_\theta  & S'\!+\!T\!-\!n^2 & (D'\!-\!E)\mathrm{c}_\theta \\
    (P'\!-\!S')\mathrm{s}_\theta\mathrm{c}_\theta & (D'\!-\!E)\mathrm{c}_\theta & S'\mathrm{c}^2_\theta\!+\!P'\mathrm{s}^2_\theta\!+\!T\!-\!n^2
    \end{array}\hspace{-4pt} \right).\hspace{6pt}
\end{eqnarray}
The degenerate matrix equation \mbox{$\mathcal{D}'\mathbf{e}=0$} is solved when the polarization angles satisfy
\begin{eqnarray}
    \label{eq:psi}
    \tan\psi&=&\frac{[(n^2-T)D'-S'E]\mathrm{c}_\theta}{(n^2-T)S' -R'L'-D'E\mathrm{c}^2_\theta},\\
    \label{eq:phi}
    \tan\phi\cos\psi&=&\frac{(P'D'-EA)\mathrm{c}_\theta}{[(n^2\!-\!P'\!-\!T)D'\!+\!(P'\!-\!S')E\mathrm{c}^2_\theta]\mathrm{s}_\theta}.
\end{eqnarray}
These polarization angles can be computed after solving $\omega$ as a function of $ck$, using which the refractive index and the dispersion symbols can be evaluated.

A numerically robust procedure for computing the unit polarization vector is by directly solving the degenerate matrix equation. 
Denoting $\mathbf{d}'_i$ the $i$-the row vector of the matrix $\mathcal{D}'$, then $\mathbf{e}\propto \alpha_1\mathbf{d}'_2\times\mathbf{d}'_3+\alpha_2\mathbf{d}'_3\times\mathbf{d}'_1+\alpha_3\mathbf{d}'_1\times\mathbf{d}'_2$, where $\alpha_i$ is an arbitrary parameter. Since $\mathcal{D}'$ is a rank-2 matrix when the dispersion relation is satisfied, the three vectors on the RHS are parallel, and at most two of them can be simultaneously zero when the plasma is magnetized. 
By summing up the three terms, it is guaranteed that the RHS is never a zero vector as long as special values of $\alpha_i$ are avoided. Then, the unit polarization vector can be determined after normalization.
Having obtained $\mathbf{e}$ in Cartesian coordinate, the polarization angles in spherical coordinate can be easily determined. An example is shown in Fig.~\ref{fig:wave}b, where $\phi$ (solid lines) and $\psi$ (dashed lines) are plotted as functions of wave frequency.

While the above procedure is generally applicable, it is instructive to note two special propagation angles.  
When $\theta=0^\circ$ or $90^\circ$, Eq.~(\ref{eq:psi}) and (\ref{eq:phi}) become indeterminate, even though the polarization vector remains well defined.
When $\theta=0^\circ$, the longitudinal electrostatic modes, which satisfy $P=0$, decouple with the transverse modes. One set of transverse modes satisfy $n^2=R$ and are right-handed (R) circularly polarized with $\tan\psi=1$; the other set of transverse modes satisfy $n^2=L$ and are left-handed (L) circularly polarized with $\tan\psi=-1$.
When $\theta=90^\circ$, the ordinary (O) wave decouples. The O wave is unmagnetized EM wave, which satisfies $n^2=P$ with the wave electric field along $\mathbf{B}_0$. The remaining modes are the extraordinary (X) wave hybridized with plasma waves, which satisfies $n^2=RL/S+T$ with $\tan\phi=S/D$ neither transverse nor longitudinal. For these modes $\cos\psi=0$, and the wave electric field is always perpendicular to the background magnetic field.

\subsection{Energy of linear waves \label{sec:energy}}
Inherited from the nonlinear fluid-Maxwell equations, the linear system also conserves energy locally. The energy density of the linear system is of $\lambda^2$ order:
\begin{equation}
    \label{eq:U2}
    U_2=\frac{\epsilon_0\mathbf{E}_1^2}{2}+\frac{\mathbf{B}_1^2}{2\mu_0}+\sum_s\frac{1}{2}\Big(m_sn_{s0}\mathbf{v}_{s1}^2+\frac{\varepsilon_s n_{s1}^2}{n_{s0}}\Big),
\end{equation}
where the last term comes from $p_{s2}$ as will become clear later when I discuss $\lambda^2$-order equations. 
The energy flux of the linear system is
\begin{eqnarray}
    \label{eq:S2}
    \mathbf{S}_2=\frac{1}{\mu_0}\mathbf{E}_1\times\mathbf{B}_1+\sum_s\varepsilon_s n_{s1}\mathbf{v}_{s1},
\end{eqnarray}
where the last term is due to thermal flux and kinetic flux does not contribute at $\lambda^2$ order. Using the first-order equations (\ref{eq:n1})-(\ref{eq:E1}), it is a straightforward calculation to show the energy conservation law on fast scales:
\begin{equation}
    \label{eq:conserve2}
    \partial_{t_0} U_2+\nabla_0\cdot\mathbf{S}_2=0.
\end{equation}
Notice that the conservation law mixes all linear waves that are present in the system. In other words, not only do linear waves contribute individually, but their interference also contributes to the total energy.

Now let us focus on the energy of a single linear wave. Using the Fourier expansion $\mathbf{E}_1=(\mathbfcal{E}e^{i\theta}+\mathbfcal{E}^*e^{-i\theta})/2$, $\mathbf{E}_1^2=(\mathbfcal{E}^2e^{2i\theta}+2\mathbfcal{E}\mathbfcal{E}^*+\mathbfcal{E}^{*2}e^{-2i\theta})/4$. We see the electric energy has a rapidly-oscillating part and a slowly-varying part. Only the later remains after averaging on $t_0$ and $\mathbf{x}_0$ scales, namely, $\langle \mathbf{E}_1^2\rangle_0= \mathbfcal{E}\mathbfcal{E}^*/2$. Following similar arguments, the averaged magnetic energy can be computed using Fourier expansion Eq.~(\ref{eq:B1wave}). Then, the averaged field energy density is $\epsilon_0\mathcal{E}_i^\dagger(2\delta^{ij}-\sum_s\omega_{ps}^2\hat{\mathbb{F}}_s^{ij}/\omega^2)\mathcal{E}_j/4$,
where I have used $\mathbb{D}_\mathbf{k} \mathbfcal{E}_{1,\mathbf{k}}=\mathbf{0}$ and Eq.~(\ref{eq:Dtensor}). Similarly, the averaged kinetic energy density is $\epsilon_0\sum_s\omega_{ps}^2 \mathcal{E}_i^\dagger(\hat{\mathbb{F}}^2_s)^{ij}\mathcal{E}_j/4\omega^2$, where I have used the self-adjoint property of $\hat{\mathbb{F}}_s$.
Finally, the averaged thermal energy density is $\epsilon_0\sum_s \omega_{ps}^2 u_s^2 \mathcal{E}_i^\dagger\hat{\mathbb{F}}_s^{ia}k_ak_b\hat{\mathbb{F}}_s^{bj}\mathcal{E}_j/4\omega^4$.
Summing up the three terms, the total energy density of the linear wave can be written as
\begin{equation}
    \label{eq:U2avg}
    \langle U_{2}\rangle_0=\frac{\epsilon_0}{4}\mathcal{E}_i^\dagger\mathbb{H}^{ij}\mathcal{E}_j.
\end{equation}
The Hamiltonian $\mathbb{H}$ of the linear wave is related to the dielectric tensor $\epsilon=\mathbb{I}+\sum_s\chi_s$ by the usual relation $\omega\mathbb{H}=\partial(\omega^2\epsilon)/\partial\omega$, where the partial derivative is at fixed $\mathbf{k}$. Since the susceptibility $\chi_s$ is related to $\hat{\mathbb{F}}_s$ by Eq.~(\ref{eq:chi}), the wave energy operator
\begin{equation}
    \label{eq:H}
    \mathbb{H}=\frac{1}{\omega}\frac{\partial\mathbb{D}}{\partial\omega}=2\mathbb{I}-\sum_s \frac{\omega_{ps}^2}{\omega} \frac{\partial\hat{\mathbb{F}}_s}{\partial\omega} ,
\end{equation}
where $\mathbb{I}$ is the idensity operator and the partial derivative is again at fixed $\mathbf{k}$.
To see the connection between Eqs.~(\ref{eq:U2avg}) and (\ref{eq:H}), note the following identity
\begin{equation}
    \label{eq:dwFhat}
    \omega \frac{\partial\hat{\mathbb{F}}_{ij}}{\partial\omega}=\hat{\mathbb{F}}_{ij} -\hat{\mathbb{F}}^2_{ij} -\frac{u^2}{\omega^2}\hat{\mathbb{F}}_{ia}k^ak^b\hat{\mathbb{F}}_{bj}.
\end{equation}
An easy way to show the above identity is to take patial derivative on both sides of $\hat{\mathbb{F}}\hat{\mathbb{F}}^{-1}=\mathbb{I}$, then $\partial\hat{\mathbb{F}}/\partial\omega=-\hat{\mathbb{F}} (\partial\hat{\mathbb{F}}^{-1}/\partial\omega)\hat{\mathbb{F}}$. Using the expression of the inverse operator, it is easy to see $\omega(\partial\hat{\mathbb{F}}^{-1}/\partial\omega)\mathbf{Z}=i\beta\mathbf{Z}\times\mathbf{b}+ 2u^2\mathbf{k}(\mathbf{k}\cdot\mathbf{Z})/\omega^2$. Replacing $\mathbf{Z}$ by $\hat{\mathbb{F}}\mathbf{Z}$, and using the Eq.~(\ref{eq:FhatZ}), the above identity is then obvious.

The averaged wave energy depends on the wave envelope.
To leading order, the wave envelope is a function of $\mathbf{x}_1-\mathbf{v}_g t_1$, where $v_g^i=\partial\omega/\partial k_i$ is the group velocity of the linear wave. Consequently, the averaged energy satisfies the advection equation on slow scales:
\begin{equation}
    \label{eq:advect}
    \partial_{t_1}\langle U_{2}\rangle_0+\mathbf{v}_g\cdot\nabla_1\langle U_{2}\rangle_0=0.
\end{equation}
This equation is consistent with the $\lambda^3$-order conservation law if and only if
\begin{equation}
    \label{eq:S2avg}
    \langle \mathbf{S}_{2}\rangle_0=\mathbf{v}_g\langle U_{2}\rangle_0.
\end{equation}
Now let me show this is indeed the case by direct calculations. The averaged Poynting flux is $\epsilon_0c^2 \mathcal{E}_a^\dagger(2k^i\delta^{ab}-k^a\delta^{ib}-k^b\delta^{ia})\mathcal{E}_b/4\omega$, and the averaged thermal flux is $\epsilon_0  \sum_s\omega_{ps}^2\mathcal{E}_a^\dagger(\partial \hat{\mathbb{F}}_s^{ab}/\partial k_i)\mathcal{E}_b/4\omega$.
For the thermal flux, I have used the identity that the partial derivative of $\hat{\mathbb{F}}$ at fixed $\omega$ is given by
\begin{equation}
    \label{eq:dkFhat}
    \frac{\partial\hat{\mathbb{F}}^{ab}}{\partial k_i}=\frac{u^2}{\omega^2} k_l(\hat{\mathbb{F}}^{ai}\hat{\mathbb{F}}^{lb} + \hat{\mathbb{F}}^{al}\hat{\mathbb{F}}^{ib}),
\end{equation}
which can be shown similarly to Eq.~(\ref{eq:dwFhat}). Summing the Poynting and the thermal fluxes, Eq.~(\ref{eq:S2avg}) is satisfied whenever $\omega v_g^i \mathcal{E}_a^*\mathbb{H}^{ab}\mathcal{E}_b=\mathcal{E}_a^*[c^2(2k^i\delta^{ab}-k^a\delta^{ib}-k^b\delta^{ia})+\sum_s\omega_{ps}^2\partial\hat{\mathbb{F}}_s^{ab}/\partial k_i]\mathcal{E}_b.$
This equation is nothing other than $\mathcal{E}_a^*d_{k_i}\mathbb{D}^{ab}\mathcal{E}_b=0$, which is trivially satisfied as a consequence of $\mathbb{D}_\mathbf{k} \mathbfcal{E}_{1,\mathbf{k}}=\mathbf{0}$ for all linear eigenmodes, whose frequency satisfies the dispersion relation and polarization solves the first-order electric-field equation. I have thus verified that the envelope of a single linear wave advects at the wave group velocity as expected. 

\section{Magnetized three-wave interactions \label{sec:three}}
Building upon a thorough understanding of linear waves, we are now ready to study their interactions. Due to these interactions, waves become coupled. Consequently, instead of passing through each other uneventfully with only linear superpositions, waves now actually ``collide" and exchange energy and momentum. In this section, I will investigate couplings mediated by three-wave interactions. These lowest-order interactions are usually the strongest whenever resonance conditions can be satisfied.

\subsection{Second-order equations\label{sec:second}}
The $\lambda^2$-order fluid-Maxwell's equations (\ref{eq:n2})-(\ref{eq:E2}) are linear partial differential equations for $\mathbf{E}_2$, $\mathbf{B}_2$, $p_{s2}$, $n_{s2}$, and  $\mathbf{v}_{s2}$ with source terms. The general solution is again a superposition of plane waves, whose spectrum is completely determined by existing linear waves in the system. 

Let us express all second-order fluctuations in terms of electric-field fluctuations:
\begin{equation}
    \label{eq:E2wave}
    \mathbf{E}_2=\frac{1}{2}\sum_{\mathbf{k}\in\mathbb{K}_2}\mathbfcal{E}_{2,\mathbf{k}} e^{i\theta_{\mathbf{k}}},
\end{equation}
where the $\lambda^2$-order spectrum $\mathbb{K}_2$ and amplitudes $\mathbfcal{E}_{2,\mathbf{k}}$ will be determined later. Using the second-order Faraday's law [Eq.~(\ref{eq:B2})], the $\lambda^2$-order magnetic field is
\begin{eqnarray}
    \label{eq:B2wave}
    \nonumber
    \mathbf{B}_2&=&\frac{1}{2}\sum_{\mathbf{k}\in\mathbb{K}_2}\frac{\mathbf{k}\times\mathbfcal{E}_{2,\mathbf{k}}}{\omega_\mathbf{k}} e^{i\theta_{\mathbf{k}}}\\
    &+&\frac{1}{2}\sum_{\mathbf{p}\in\mathbb{K}_1}\Big(\frac{\nabla_1\times\mathbfcal{E}_{1,\mathbf{p}}}{i\omega_\mathbf{p}}+ \frac{\mathbf{p}\times\partial_{t_1}\mathbfcal{E}_{1,\mathbf{p}}}{i\omega_\mathbf{p}^2}\Big)e^{i\theta_{\mathbf{p}}}.
\end{eqnarray}
The second line involves slow derivatives of linear wave amplitudes, which are unknown at this point.

To solve the fluid equations, let us first express $p_{s2}$ in terms of $n_{s2}$. Using $p_{s1}=\varepsilon_s n_{s1}$ [Eq.~(\ref{eq:p1wave})], many terms in the second-order pressure equation cancels. Integrating Eq.~(\ref{eq:p2}) on $t_0$ time scale, the second-order pressure is
\begin{equation}
    \label{eq:p2wave}
    p_{s2}=\varepsilon_s\Big(n_{s2}+\frac{\xi_s-1}{2}\frac{n_{s1}^2}{n_{s0}}\Big).
\end{equation}
Notice that due to quadratic nonlinearities, the average $\langle p_{s2}\rangle_0$ is in general nonzero. We see that $p_{s2}$ is related to $n_{s1}^2$, as anticipated from the energy of linear waves.

Next, let us express $n_{s2}$ in terms of $\mathbf{v}_{s2}$. Suppose the Fourier expansion of the second-order velocity is  $\mathbf{v}_{s2}=\sum_\mathbf{k}\exp(i\theta_\mathbf{k})\mathbfcal{V}_{s2,\mathbf{k}}/2$. Then, substituting $\mathbf{v}_{s1}$ and $n_{s1}$ [Eqs.~(\ref{eq:v1wave}) and (\ref{eq:n1wave})] into the second-order continuity equation [Eq.~(\ref{eq:n2})], the $\lambda^2$-order density is
\begin{eqnarray}
    \label{eq:n2wave}
    \nonumber
    \frac{n_{s2}}{n_{s0}}&=&\frac{1}{2}\sum_{\mathbf{k}}\frac{\mathbf{k}\cdot\mathbfcal{V}_{s2,\mathbf{k}}}{\omega_\mathbf{k}} e^{i\theta_\mathbf{k}}\\
    \nonumber
    &+&\frac{e_s}{2m_s}\sum_{\mathbf{p}\in\mathbb{K}_1}\Big(\frac{\mathbf{p}\cdot\hat{\mathbb{F}}_{s,\mathbf{p}}\partial_{t_1}\mathbfcal{E}_{1,\mathbf{p}}}{\omega_\mathbf{p}^3}+\frac{\nabla_1\cdot\hat{\mathbb{F}}_{s,\mathbf{p}}\mathbfcal{E}_{1,\mathbf{p}}}{\omega_\mathbf{p}^2}\Big) e^{i\theta_\mathbf{p}}\\
    &-&\frac{e_s^2}{4m_s^2} \sum_{\mathbf{p},\mathbf{q}\in\mathbb{K}_1} \frac{(\mathbf{p}+\mathbf{q})\cdot\mathbf{C}^s_{\mathbf{p},\mathbf{q}}}{(\omega_\mathbf{p}+\omega_\mathbf{q})\omega_\mathbf{q}} e^{i\theta_\mathbf{p}+i\theta_\mathbf{q}},
\end{eqnarray}
where summation on the first line is over the spectrum of $\mathbf{v}_{s2}$.
Since thermal effect does not enter through the continuity equation directly, the above expression is identical to the cold-fluid case, where the current beating is
\begin{equation}
    \label{eq:Cbeat}
    \mathbf{C}^s_{\mathbf{p},\mathbf{q}}=\frac{(\hat{\mathbb{F}}_{s,\mathbf{p}}\mathbfcal{E}_{1,\mathbf{p}}) (\mathbf{q}\cdot\hat{\mathbb{F}}_{s,\mathbf{q}}\mathbfcal{E}_{1,\mathbf{q}})}{\omega_\mathbf{p}\omega_\mathbf{q}}.
\end{equation}
The current beating comes from the nonlinearty $n_{s1}\mathbf{v}_{s1}$, and thermal effects enter only indirectly through the warm forcing operator.

Eliminating $p_{s2}$ and $n_{s2}$, we can now solve for $\mathbf{v}_{s2}$. 
Using Eq.~(\ref{eq:v1}) to simplify Eq.~(\ref{eq:v2}), the equation 
is of the form $\sum_\mathbf{k}(\mathbfcal{V}_{2,\mathbf{k}}-i\beta \mathbfcal{V}_{2,\mathbf{k}}\times\mathbf{b}-u^2\mathbf{k}\mathbf{k}\cdot\mathbfcal{V}_{2,\mathbf{k}}/\omega_\mathbf{k}^2)\exp(i\theta_\mathbf{k})=\sum_\mathbf{k}\mathbf{Z}_\mathbf{k}\exp(i\theta_\mathbf{k})$, where I have suppressed the species index for simplicity. Then, using Eq.~(\ref{eq:FhatZ}) of the warm forcing operator, the solution is
\begin{eqnarray}
    \label{eq:v2wave}
    \mathbf{v}_{2}&=&\frac{ie}{2m}\sum_{\mathbf{k}\in\mathbb{K}_2}\frac{\hat{\mathbb{F}}_{\mathbf{k}}\mathbfcal{E}_{2,\mathbf{k}}}{\omega_\mathbf{k}} e^{i\theta_\mathbf{k}} \\
    \nonumber
    &+&\!\frac{e}{2m}\hspace{-5pt}\sum_{\mathbf{p}\in\mathbb{K}_1}\hspace{-5pt} \frac{\hat{\mathbb{F}}_{\mathbf{p}}}{\omega_\mathbf{p}^2}\! \Big[\!\Big(\!\mathbb{I}\!+\!\frac{u^2\!\mathbf{p}\mathbf{p}}{\omega_\mathbf{p}^2}\!\Big)\!\partial_{t_1}\hspace{-5pt}+\!\frac{u^2\!(\!\mathbf{p}\!\nabla_1\!+\!\nabla_1 \!\mathbf{p}\!)}{\omega_\mathbf{p}}\!\Big]\hat{\mathbb{F}}_{\mathbf{p}}\mathbfcal{E}_{1,\mathbf{p}} e^{i\theta_\mathbf{p}}    \\
    \nonumber
    &-&\frac{e^2}{4m^2}\sum_{\mathbf{p},\mathbf{q}\in\mathbb{K}_1}\frac{\hat{\mathbb{F}}_{\mathbf{p}+\mathbf{q}}(\mathbf{L}_{\mathbf{p},\mathbf{q}}+\mathbf{T}_{\mathbf{p},\mathbf{q}}+\mathbf{U}_{\mathbf{p},\mathbf{q}})}{\omega_\mathbf{p}+\omega_\mathbf{q}} e^{i\theta_\mathbf{p}+i\theta_\mathbf{q}}.
\end{eqnarray}
The spectrum $\mathbfcal{V}_{2,\mathbf{k}}$ can now be read out from the above equation, and explicit expressions of $n_{s2}$ and $p_{s2}$ can then be obtained. 
On the third line of Eq.~(\ref{eq:v2wave}), the first term $\mathbf{L}_{\mathbf{p},\mathbf{q}}$ is the longitudinal beating introduced by the $\mathbf{v}_1\times\mathbf{B}_1$ nonlinearity:
\begin{equation}
    \label{eq:Lbeat}
    \mathbf{L}^s_{\mathbf{p},\mathbf{q}}=\frac{(\hat{\mathbb{F}}_{s,\mathbf{p}}\mathbfcal{E}_{1,\mathbf{p}})\times (\mathbf{q}\times\mathbfcal{E}_{1,\mathbf{q}})}{\omega_\mathbf{p}\omega_\mathbf{q}}.
\end{equation}
The second term $\mathbf{T}_{\mathbf{p},\mathbf{q}}$ is the turbulent beating introduced by the \mbox{$\mathbf{v}_1\cdot\nabla_0\mathbf{v}_1$} nonlinearity:
\begin{equation}
    \label{eq:Tbeat}
    \mathbf{T}^s_{\mathbf{p},\mathbf{q}}=\frac{(\hat{\mathbb{F}}_{s,\mathbf{p}}\mathbfcal{E}_{1,\mathbf{p}}) (\mathbf{p}\cdot\hat{\mathbb{F}}_{s,\mathbf{q}}\mathbfcal{E}_{1,\mathbf{q}})}{\omega_\mathbf{p}\omega_\mathbf{q}}.
\end{equation}
These two terms are the same as in the cold case, except that the cold forcing operator is now replaced by the warm forcing operator. Additionally, the thermal effect introduces thermal beating as a third term:
\begin{equation}
    \label{eq:Ubeat}
    \mathbf{U}^s_{\mathbf{p},\mathbf{q}}=\frac{u_s^2}{\omega_\mathbf{p}\omega_\mathbf{q}}\Big[\frac{(\mathbf{p}+\mathbf{q})(\mathbf{p}+\mathbf{q})}{1+\omega_\mathbf{q}/\omega_\mathbf{p}}+(\xi_s-2)\mathbf{p}\mathbf{p}\Big]\cdot \mathbf{C}^s_{\mathbf{p},\mathbf{q}},
\end{equation}
which is caused by nonlinearities in $n_{s2}$ and $p_{s2}$. The turbulent and thermal beatings can be rewritten in terms of the velocity perturbation $\mathbfcal{V}_{1,\mathbf{k}}$. These beatings are purely fluid effects, which exist even when the fluid is neutral (Appendix~\ref{app:neutral}). On the other hand, the longitudinal and current beatings are genuine electromagnetic nonlinearities, whereby transverse EM waves in the vacuum become mixed with the otherwise longitudinal motion of the plasma.

Having expressed all fluctuations in terms of electric-field fluctuations, we can now solve for the electric field. Substituting in $\mathbf{B}_2$ and $\mathbf{v}_{s2}$ into Eq.~(\ref{eq:E2}), the $\lambda^2$-order electric-field equation can be grouped into four sets of terms, involving $\mathbfcal{E}_{2,\mathbf{k}}$, $\partial_{t_1}\mathbfcal{E}_{1,\mathbf{k}}$, $\nabla_1 \mathbfcal{E}_{1,\mathbf{k}}$, and $\mathbfcal{E}_{1,\mathbf{p}}\mathbfcal{E}_{1,\mathbf{q}}$.
Differentiating on $t_0$ scale, the first set simplifies to $\mathbb{D}_\mathbf{k} \mathbfcal{E}_{2,\mathbf{k}}$ using Eq.~(\ref{eq:Dtensor}).
The second set simplifies to 
$(\partial\mathbb{D}_\mathbf{k}/\partial\omega_\mathbf{k})\partial_{t_1}\mathbfcal{E}_{1,\mathbf{k}}$, using $\mathbb{D}_\mathbf{k} \mathbfcal{E}_{1,\mathbf{k}}=\mathbf{0}$ and Eq.~(\ref{eq:dwFhat}).
The third set simplifies to
$-(\partial \mathbb{D}_\mathbf{k}/\partial\mathbf{k})\cdot\nabla_1\mathbfcal{E}_{1,\mathbf{k}}$, using Eq.~(\ref{eq:dkFhat}).
Finally, the fourth set encapsulates all beatings. With all these simplifications, the second-order electric-field equation is then
\begin{eqnarray}
    \label{eq:E2eq}
    \nonumber
    &&\sum_{\mathbf{k}\in\mathbb{K}_2}\mathbb{D}_\mathbf{k} \mathbfcal{E}_{2,\mathbf{k}} e^{i\theta_\mathbf{k}}\\
    &+&i\sum_{\mathbf{k}\in\mathbb{K}_1}
    \Big(\frac{\partial\mathbb{D}_\mathbf{k}}{\partial\omega_\mathbf{k}}\partial_{t_1} 
    -\frac{\partial \mathbb{D}_\mathbf{k}}{\partial\mathbf{k}}\cdot\nabla_1\Big)\mathbfcal{E}_{1,\mathbf{k}}e^{i\theta_\mathbf{k}} \\
    \nonumber
    &=&\frac{i}{2}\sum_{\mathbf{p},\mathbf{q}\in\mathbb{K}_1}\mathbf{S}_{\mathbf{p},\mathbf{q}} e^{i\theta_\mathbf{p}+i\theta_\mathbf{q}} ,
\end{eqnarray}
where the scattering strength summed over all plasma species is
\begin{equation}
    \label{eq:Spq}
    \mathbf{S}_{\mathbf{p},\mathbf{q}}=\sum_s\frac{e_s\omega^2_{ps}}{2m_s}\Big(\mathbf{R}^s_{\mathbf{p},\mathbf{q}}+\mathbf{R}^s_{\mathbf{q},\mathbf{p}}\Big).
\end{equation}
The above equation is formally identical to the cold case, except that the dispersion tensor $\mathbb{D}$ now contains thermal modifications. 
In addition, thermal effects directly enter the quadratic response of each species:
\begin{equation}
    \label{eq:Rpq}
    \mathbf{R}^s_{\mathbf{p},\mathbf{q}}=\hat{\mathbb{F}}_{s,\mathbf{p}+\mathbf{q}}\big(\mathbf{L}^s_{\mathbf{p},\mathbf{q}}+\mathbf{T}^s_{\mathbf{p},\mathbf{q}}+\mathbf{U}^s_{\mathbf{p},\mathbf{q}}\big)+\big(1+\frac{\omega_p}{\omega_q}\big) \mathbf{C}^s_{\mathbf{p},\mathbf{q}}.
\end{equation}
The thermal beating $\mathbf{U}^s_{\mathbf{p},\mathbf{q}}$ vanishes when the thermal speed $u_s^2\rightarrow 0$, while the three other beatings remain finite when the species becomes cold.

Since the second-order electric-field equation must be satisfied for each Fourier component, the equation can be split into two sets. The first set involves only $\lambda$-order spectrum $\mathbb{K}_1$. Suppose within the spectral bandwidths, $\mathbf{k}, \mathbf{p}, \mathbf{q} \in\mathbb{K}_1$ satisfy the three-wave resonance conditions $\mathbf{k}=\mathbf{p}+\mathbf{q}$ and $\omega_\mathbf{k}=\omega_\mathbf{p}+\omega_\mathbf{q}$, then the on-shell equation is of the form
\begin{equation}
    \label{eq:onShell}
    \Big(\frac{\partial\mathbb{D}_\mathbf{k}}{\partial\omega_\mathbf{k}}\partial_{t_1} 
    -\frac{\partial \mathbb{D}_\mathbf{k}}{\partial\mathbf{k}}\cdot\nabla_1\Big)\mathbfcal{E}_{1,\mathbf{k}}=\mathbf{S}_{\mathbf{p},\mathbf{q}}.
\end{equation}
The RHS is simply zero if $\mathbf{k}\in\mathbb{K}_1$ is not in resonance with other waves. 
The second set of equations generate $\mathbb{K}_2$ from $\mathbb{K}_1$. To aviod secular behavior, we can demand that \mbox{$\mathbb{K}_2\cap\mathbb{K}_1$} is the empty set, so that all $\mathbf{k}\in\mathbb{K}_2$ is generated by off-resonance beatings. Then, each off-shell equation is of the form
\begin{equation}
    \label{eq:offShell}
    \mathbb{D}_\mathbf{k} \mathbfcal{E}_{2,\mathbf{k}}=i \mathbf{S}_{\mathbf{p},\mathbf{q}},
\end{equation}
where $\mathbf{k}=\mathbf{p}+\mathbf{q}$ and $\omega_\mathbf{k}=\omega_\mathbf{p}+\omega_\mathbf{q}$ are still satisfied, except that $\mathbf{k}\notin\mathbb{K}_1$ is no longer a linear eigenmode. In other words, the wave dispersion relation is not satisfied for off-shell waves, so the operator $\mathbb{D}_\mathbf{k}$ can be inverted to give $\mathbfcal{E}_{2,\mathbf{k}}=i \mathbb{D}_\mathbf{k}^{-1} \mathbf{S}_{\mathbf{p},\mathbf{q}}$. Thus, the $\lambda^2$-order spectrum is completely determined by the $\lambda$-order spectrum.
After solving both the on-shell and the off-shell equations, $\mathbf{B}_2, \mathbf{v}_{s2}, n_{s2}, p_{s2}$ can be determined and the second-order equations are then solved.

Now that the $\lambda^2$-order equations have been formally solved, it is important to note $\mathbf{S}_{\mathbf{p},\mathbf{q}}$ satisfies a number of identities, which are required in order for the solutions to be valid. 
First, from its definition, it is obvious that $\mathbf{S}_{\mathbf{p},\mathbf{q}}=\mathbf{S}_{\mathbf{q},\mathbf{p}}$ is symmetric, which is expected from the symmetry of three-wave interactions.
Second, it is a straightforward calculation to check that $\mathbf{S}_{\mathbf{p},\mathbf{q}}^*=-\mathbf{S}_{-\mathbf{q},-\mathbf{p}}$ satisfies the reality condition. Consequently, all second-order fluctuations are real-valued. 
Finally, $\mathbf{S}_{\mathbf{p},-\mathbf{p}}=\mathbf{0}$ is secular-free. In other words, a wave does not beat with itself to generate a mode with both $\omega=0$ and $\mathbf{k}=\mathbf{0}$. 
To show the last identity, notice that for each species $\mathbf{U}_{\mathbf{p},-\mathbf{p}}+\mathbf{U}_{-\mathbf{p},\mathbf{p}} =\mathbf{0}$. 
Moreover, since $\omega_{-\mathbf{p}}=-\omega_{\mathbf{p}}$, we have $\mathbf{R}_{\mathbf{p},-\mathbf{p}}+\mathbf{R}_{-\mathbf{p},\mathbf{p}}=\hat{\mathbb{F}}_\mathbf{0}(\mathbf{L}_{\mathbf{p},-\mathbf{p}}+\mathbf{L}_{-\mathbf{p},\mathbf{p}}+\mathbf{T}_{\mathbf{p},-\mathbf{p}}+\mathbf{T}_{-\mathbf{p},\mathbf{p}})=-i\beta_\mathbf{p}\hat{\mathbb{F}}_\mathbf{0}[(\mathbf{p}\cdot \hat{\mathbb{F}}_\mathbf{p}^*\mathbfcal{E}_\mathbf{p}^*)(\hat{\mathbb{F}}_\mathbf{p}\mathbfcal{E}_\mathbf{p}\times\mathbf{b})+\text{c.c.}]/\omega_\mathbf{p}^2$. 
To see what $\hat{\mathbb{F}}_\mathbf{0}$ is, we can take limits $\omega\rightarrow 0$ and $\mathbf{k}\rightarrow\mathbf{0}$. Although $\hat{\mathbb{F}}_\mathbf{0}$ depends on the how these two limits are taken, $\mathbf{R}_{\mathbf{p},-\mathbf{p}}+\mathbf{R}_{-\mathbf{p},\mathbf{p}}=\mathbf{0}$ is independent of the limiting procedure. Therefore, $\mathbf{S}_{\mathbf{p},-\mathbf{p}}=\mathbf{0}$ is always satisfied.

\subsection{On-shell equations and action conservation\label{sec:shell}}
While the off-shell equations are easy to solve, the on-shell equations are where the nontrivial dynamics is contained. These equations are nonlinearly-coupled advection equations. 
Due to the vector nature of these equations, not only does wave amplitude change, but the wave polarization can also evolve. Moreover, the wave phase, trajectory, and angular momentum are usually altered as well due to three-wave interactions.

Since the wave polarization can change in general, it is important to verify whether the on-shell equation is compatible with the first-order equation. Suppose $\mathbb{D}\mathbfcal{E}=\mathbf{0}$ is satisfied over the entire space before the waves encounter, then $\mathbb{D}\mathbfcal{E}=\mathbf{0}$ will be satisfied for all time if $\mathbb{D}\partial_{t_1}\mathbfcal{E}=\mathbf{0}$ for all $\mathbf{x}_1$. Using the on-shell equation (\ref{eq:onShell}), this compatibility condition is satisfied if
\begin{equation}
    \label{eq:polCondition}
    \mathbb{D}\Big(\frac{\partial\mathbb{D}}{\partial\omega}\Big)^{-1}\Big(\frac{\partial\mathbb{D}}{\partial\mathbf{k}}\cdot\nabla_1\mathbfcal{E}+\mathbf{S}\Big)=\mathbf{0},
\end{equation}
where I have used the fact that the Hamiltonian $\omega\mathbb{H}=\partial\mathbb{D}/\partial\omega$ is invertible.
Since the dispersion operator $\mathbb{D}$ is degenerate for linear eigenmodes, the above condition only requires that $(\partial\mathbb{D}/\partial\mathbf{k})\cdot\nabla_1\mathbfcal{E}+\mathbf{S}$ is in the null space of $\mathbb{D}(\partial\mathbb{D}/\partial\omega)^{-1}$.
Notice that the rank of $\mathbb{D}$ is at most two, so the above condition imposes at most two constraints. Therefore, there always remains degree of freedom allowing $\mathbfcal{E}$ to evolve in time.

The compatibility condition can be used to remove the redundant degree of freedom of the on-shell equation. Taking total $\mathbf{k}$ derivative on both sides of $\mathbb{D}\mathbfcal{E}=\mathbf{0}$, where $\omega$ and $\mathbfcal{E}$ are now regarded as functions of $\mathbf{k}$, we have $(d\mathbb{D}/d\mathbf{k})\mathbfcal{E}+\mathbb{D}d\mathbfcal{E}/d\mathbf{k}=\mathbf{0}$.
Here, the total derivative of the dispersion tensor is 
$d\mathbb{D}/d\mathbf{k}=\mathbf{v}_g \partial\mathbb{D}/\partial\omega+\partial\mathbb{D}/\partial\mathbf{k}$, where $\mathbf{v}_g=\partial\omega/\partial\mathbf{k}$ is the wave group velocity.
Using the wave energy operator [Eq.~(\ref{eq:H})], the on-shell equation (\ref{eq:onShell}) becomes $\omega\mathbb{H}(\partial_{t_1}+\mathbf{v}_g\cdot\nabla_1)\mathbfcal{E}+\mathbb{D}\partial_l d\mathbfcal{E}/dk_l=\mathbf{S}$.
The compatibility condition Eq.~(\ref{eq:polCondition}) then guarantees that the advection keeps $\mathbfcal{E}$ inside the eigenspace. To be more specific, denoting $\Pi$ the projection operator into the null space of $\mathbb{D}$, then after applying the compatibility condition, the on-shell equation is reduced to
\begin{equation}
    \label{eq:onShell_proj}
    \omega\mathbb{H}(\partial_{t_1}+\mathbf{v}_g\cdot\nabla_1)\mathbfcal{E}=\mathbf{S}^\pi,
\end{equation}
where $\mathbf{S}^\pi=\mathbb{H}\,\Pi\,\mathbb{H}^{-1}\mathbf{S}$ is the eigen projection of $\mathbf{S}$. This somewhat abstract notation is illustrated using cold unmagnetized plasma as an example in Appendix~\ref{app:onShell}.
Since $\mathbfcal{E}^\dagger\mathbb{D}=\mathbf{0}$ for eigenmodes, $\mathbfcal{E}^\dagger\mathbf{S}^\pi=\mathbfcal{E}^\dagger\mathbf{S}$. We see only the eigen projection of the scattering strength affects the evolution of the linear wave.

Now let us focus on the simplest nontrivial case, namely, the resonant interaction between three on-shell waves. Without loss of generality, the resonance conditions can be written as 
\begin{eqnarray}
    \label{eq:kResonance}
    \mathbf{k}_1=\mathbf{k}_2+\mathbf{k}_3,\\
    \label{eq:wResonance}
    \omega_1=\omega_2+\omega_3,
\end{eqnarray}
where $\omega_i$ are positive. For simplicity, I will abbreviate $\mathbfcal{E}_{j}:=\mathbfcal{E}_{1,\mathbf{k}_j}$. Moreover, since the slow dynamics is on $t_1$ and $\mathbf{x}_1$ scales only, I will suppress the index of the temporal and spatial scales. Then, the three on-shell equations can be written as
\begin{eqnarray}
    \label{eq:dE1}
    \omega_1\mathbb{H}_1d \mathbfcal{E}_1=\mathbf{S}_{2,3}^\pi,\\
    \label{eq:dE2}
    \omega_2\mathbb{H}_2 d \mathbfcal{E}_2=\mathbf{S}_{\bar{3},1}^\pi,\\
    \label{eq:dE3}
    \omega_3\mathbb{H}_3 d \mathbfcal{E}_3=\mathbf{S}_{1,\bar{2}}^\pi,
\end{eqnarray}
where $\bar{j}:=-j$, and $d:=\partial_t+\mathbf{v}_g \cdot \nabla$ is the convective derivative at respective wave group velocities.
Notice that the group velocity $\mathbf{v}_g$ is in general not aligned with $\mathbf{k}$ when the plasma is magnetized.

What is nontrivial about the on-shell equations is that they guarantee action conservation for resonant three-wave interactions. The conservation laws are consequences of the action identity
\begin{equation}
    \label{eq:actionID}
    \frac{\mathbfcal{E}_1\cdot \mathbf{S}_{\bar{2},\bar{3}}}{\omega_1^2}=\frac{\mathbfcal{E}_2^*\cdot \mathbf{S}_{\bar{3},1}}{\omega_2^2}=\frac{\mathbfcal{E}_3^*\cdot \mathbf{S}_{1,\bar{2}}}{\omega_3^2}.
\end{equation}
Using this identity, $\mathbfcal{E}^\dagger\mathbf{S}^\pi=\mathbfcal{E}^\dagger\mathbf{S}$, and $\mathbf{S}_{-\mathbf{p},-\mathbf{q}}=-\mathbf{S}^*_{\mathbf{p},\mathbf{q}}$, it is easy to show the action conservation laws
\begin{eqnarray}
    \label{eq:action}
    d\frac{\langle U_1\rangle}{\omega_1}=-d\frac{\langle U_2\rangle}{\omega_2}=-d\frac{\langle U_3\rangle}{\omega_3},
\end{eqnarray}
where $\langle U_j\rangle$ is the energy of wave $j$ averaged on fast scales [Eq.~(\ref{eq:U2avg})]. 
The action conservation laws are manifestations of the Feynman rules of three-wave interactions \cite{Shi2016effective}:  each quanta of wave ``1" is converted to a quanta of wave ``2" and a quanta of wave ``3", or \textit{vice versa}. 
Using the action conservation laws and $\omega_1=\omega_2+\omega_3$, the total wave energy is also conserved:
\begin{equation}
    \label{eq:action123}
    d\langle U_1\rangle+d\langle U_2\rangle+d\langle U_3\rangle=0.
\end{equation}
Notice that the above conservation laws hold only when the three waves are in resonance.

The action identity can be shown by direct calculations. During the calculation, one will encounter terms like $\mathbfcal{F}_1^* \cdot\mathbfcal{F}_2$, where $\mathbfcal{F}_j=\hat{\mathbb{F}}_j\mathbfcal{E}_j$. Such terms can be simplified using the following quadratic identity of the Forcing operator:
\begin{equation}
    \label{eq:Fhat1Fhat2}
    (\beta_1-\beta_2)\hat{\mathbb{F}}_1\hat{\mathbb{F}}_2=\beta_1\hat{\mathbb{F}}_1\mathbb{P}_2-\beta_2\mathbb{P}_1^\dagger\hat{\mathbb{F}}_2,
\end{equation}
which can be obtained from Eq.~(\ref{eq:Fhat}) using property of the cold forcing operator \cite{Shi2017three}: $(\beta_1-\beta_2)\mathbb{F}_1\mathbb{F}_2=\beta_1\mathbb{F}_1-\beta_2\mathbb{F}_2$.
A suite of similar identities can be obtained using $\hat{\mathbb{F}}_{\bar{j}}=\hat{\mathbb{F}}_j^*$ and $\hat{\mathbb{F}}^\dagger=\hat{\mathbb{F}}$.
The product is then
\begin{eqnarray}
    \label{eq:product}
    \nonumber
    \mathbfcal{F}_1^* \!\cdot\!\mathbfcal{F}_2&=&\frac{\omega_1}{\omega_3}(\mathbfcal{E}_1^* \!\cdot\!\mathbfcal{F}_2)-\frac{\omega_2}{\omega_3}(\mathbfcal{E}_2\!\cdot\!\mathbfcal{F}_1^*)\\
    \nonumber
    &+&\frac{u^2}{\omega_3}\Big[\frac{(\mathbf{k}_1\!\cdot\! \mathbfcal{F}_1^*)(\mathbf{k}_1\!\cdot\! \mathbfcal{F}_2)}{\omega_1}- \frac{(\mathbf{k}_2\!\cdot\! \mathbfcal{F}_1^*)(\mathbf{k}_2\!\cdot\! \mathbfcal{F}_2)}{\omega_2}\Big],
\end{eqnarray}
where I have used property of the pressure operator: $\mathbb{P}\mathbf{Z}=\mathbf{Z}+u^2(\mathbf{k}\cdot\hat{\mathbb{F}}\mathbf{Z})\mathbf{k}/\omega^2$. 
The action identity [Eq.~(\ref{eq:actionID})] can then be shown by straightforward calculation of \mbox{$\mathbfcal{E}_1\cdot(\mathbf{R}_{\bar{2},\bar{3}}+\mathbf{R}_{\bar{3},\bar{2}})/\omega_1^2$}, and comparing it with the other two terms of the same structure.

In fact, terms in the action identity can be organized into a very simple and intuitive form:
\begin{eqnarray}
    \label{eq:E1S23}
    \frac{c\mathbfcal{E}_1\cdot(\mathbf{R}^s_{\bar{2},\bar{3}}+\mathbf{R}^s_{\bar{3},\bar{2}})}{\omega_1^2}=\frac{\mathcal{E}_1\mathcal{E}_2^*\mathcal{E}_3^*}{\omega_1\omega_2\omega_3} \big(\Theta^s+\Phi^s\big).
\end{eqnarray}
Here, $\mathcal{E}_i$ is the scalar amplitude such that 
$\mathbfcal{E}_i=\mathcal{E}_i\mathbf{e}_i$, where $\mathbf{e}_i$ is the unit polarization vector.
In the above expression,
$\Theta^s$ and $\Phi^s$ are the nondimensionalized electromagnetic and the thermal scattering strengths.
The electromagnetic scattering $\Theta^s$ is due to the $P_s^i(\partial_iA_l)J^l_s$ coupling in the Lagrangian \cite{Shi2017three}, where $\mathbf{P}_s$ is the displacement and $\mathbf{J}_s$ is the current of species $s$ in response to perturbations of the gauge field $\mathbf{A}$. Same as the cold case, the electromagnetic scattering contains six permutations:
\begin{equation}
    \label{eq:Theta}
    \Theta^s=\Theta^s_{1,\bar{2}\bar{3}}+\Theta^s_{\bar{2},\bar{3}1}+\Theta^s_{\bar{3},1\bar{2}}+\Theta^s_{1,\bar{3}\bar{2}}+\Theta^s_{\bar{3},\bar{2}1}+\Theta^s_{\bar{2},1\bar{3}}.
\end{equation}
Each scattering channel, which satisfies $\Theta^s_{i,\bar{j}\bar{l}}=\Theta^{s*}_{\bar{i},jl}$ is given by the simple formula
\begin{equation}
    \label{eq:Thetaijl}
    \Theta^s_{i,jl}=\frac{1}{\omega_j}(c\mathbf{k}_i\cdot\mathbf{f}_{s,j}) (\mathbf{e}_i\cdot\mathbf{f}_{s,l}),
\end{equation}
where $\mathbf{f}_{s,j}:=\hat{\mathbb{F}}_{s,j}\mathbf{e}_j$.
The thermal scattering $\Phi^s$ is due to warm-fluid effects, which is present even in neutral fluids (Appendix~\ref{app:neutral}). The thermal scattering contains four contributions
\begin{equation}
    \label{eq:Phi}
    \Phi^s=\Phi^s_0+\Phi^s_{1}+\Phi^s_{\bar{2}}+\Phi^s_{\bar{3}}.
\end{equation}
The symmetric thermal scattering is formed by contracting $\mathbf{f}$ with its own wave vector:
\begin{equation}
    \label{eq:Phi0}
    \Phi^s_0=-\frac{(\xi_s-2)u_s^2}{c^2\omega_1\omega_2\omega_3}(c\mathbf{k}_1\!\cdot\!\mathbf{f}_{s,1})(c\mathbf{k}_2\!\cdot\!\mathbf{f}_{s,2}^*)(c\mathbf{k}_3\!\cdot\!\mathbf{f}_{s,3}^*).
\end{equation}
On the other hand, the skewed symmetric thermal scattering, which satisfies $\Phi^s_{\bar{j}}=-\Phi^{s}_j$, is formed by contracting $\mathbf{f}$ with a common wave vector:
\begin{equation}
    \label{eq:Phij}
    \Phi^s_j=-\frac{u_s^2}{c^2\omega_1\omega_2\omega_3}(c\mathbf{k}_j\!\cdot\!\mathbf{f}_{s,1})(c\mathbf{k}_j\!\cdot\!\mathbf{f}_{s,2}^*)(c\mathbf{k}_j\!\cdot\!\mathbf{f}_{s,3}^*).
\end{equation}
Since $\Phi^s$ is proportional to $u_s^2/c^2$, it is usually very small, otherwise a relativistic plasma model is required in the first place. 
It is obvious that when the species is cold, the thermal scattering vanishes and the three-wave scattering reduces to purely electromagnetic scattering in the cold-fluid case.

\subsection{Coupling coefficient and growth rate\label{sec:coupling}}
When polarization is not of concern, the on-shell equations can be reduced to scalar-amplitude equations called the three-wave equations, which contain a single essential parameter: the coupling coefficient. Denoting $\mathbfcal{E}=\mathcal{E}\mathbf{e}$, 
we can define the wave energy coefficient
\begin{equation}
    \label{eq:ucoef}
    u=\frac{1}{2}\mathbf{e}^\dagger \mathbb{H}\mathbf{e},
\end{equation}
such that the averaged wave energy $\langle U\rangle=\epsilon_0 u |\mathcal{E}|^2/2$. 
It is then convenient to normalize the scalar amplitude by
\begin{equation}
    \label{eq:a}
    a=\frac{e\mathcal{E}}{m_e c\omega} u^{1/2},
\end{equation}
where $e$ and $m_e$ are the charge and mass of electrons. With this normalization, the wave energy $\langle U\rangle\propto \omega^2 |a|^2$.
Notice that the decomposition $\mathbfcal{E}=\mathcal{E}\mathbf{e}$ is not unique, and we can always perform $U(1)$ rotations $\mathcal{E}\rightarrow \mathcal{E} e^{i\alpha}$ and $\mathbf{e}\rightarrow \mathbf{e} e^{-i\alpha}$ such that the vector amplitude $\mathbfcal{E}$ is invariant.
To remove this arbitrariness, we can impose the condition that $\mathcal{E}\in\mathbb{R}$ is real-valued. Then, the decomposition is unique up to the $\mathbb{Z}_2$ symmetry $\mathcal{E}\rightarrow -\mathcal{E}$ and $\mathbf{e}\rightarrow -\mathbf{e}$. With this reduced symmetry, the normalized scalar amplitude is also real-valued.

The equation for the normalized scalar amplitude can be derived from the on-shell equations. Allowing the polarization to evolve, the derivative $d(\mathcal{E}u^{1/2})=[\mathbf{e}^\dagger\mathbb{H} d(\mathcal{E}\mathbf{e}) +\text{c.c.}]/4u^{1/2}$. 
Then, the real-valued amplitude evolves according to $d a_1=e(\mathbf{e}_1^\dagger\mathbf{S}_{2,3}+\text{c.c.})/(4m_e c\omega_1^2u_1^{1/2})$. Using Eq.~(\ref{eq:Spq}) for $\mathbf{S}_{2,3}$ and Eq.~(\ref{eq:E1S23}) for the inner product, the normalized scalar amplitudes satisfy the following three-wave equations:
\begin{eqnarray}
    \label{eq:da1}
    d a_1&=&-\frac{\Gamma_r}{\omega_1}a_2 a_3,\\
    \label{eq:da2}
    d a_2&=&\phantom{-}\frac{\Gamma_r}{\omega_2}a_1 a_3,\\
    \label{eq:da3}
    d a_3&=&\phantom{-}\frac{\Gamma_r}{\omega_3}a_1 a_2,
\end{eqnarray}
where the convective derivatives are at respective group velocities of the three waves. Due to the $\mathbb{Z}_2$ symmetry, only the relative signs of the above equations are of importance. The essential parameter of the three-wave equations is the coupling coefficient $\Gamma_r$, which is the real part of the complex-valued coupling coefficient
\begin{equation}
    \label{eq:coupling}
    \Gamma=\sum_s\frac{Z_s\omega_{ps}^2(\Theta^s+\Phi^s)}{4M_s(u_1u_2u_3)^{1/2}}.
\end{equation}
Here, $Z_s:=e_s/e$ and $M_s:=m_s/m_e$ are the normalized charge and mass of species $s$. As a consequence of wave interference, contributions of different plasma species add up in the complex plane. Moreover, the three waves also interfere. 
The three-wave interference depends on the relative wave phase, whose change corresponds to a rotation of $\Gamma$ in the complex plane. When the three waves are phase locked, $\Gamma_r=|\Gamma|$ is maximized. In this case, the beat wave of $a_1$ and $a_2$ are in phase with $a_3$, so the plasma responses constructively interfere. 
The above formula is formally identical to the cold case, except for the extra $\Phi$ term due to thermal scattering.

The three-wave coupling may be small for three distinct reasons \cite{Shi2018laser}. First, the coupling coefficient $\Gamma$ may be interference-suppressed because terms in the summation cancel one another. In this case, although scattering due to each species is appreciable, the nonlinear responses are of opposite phases and destructively interfere. Second, $\Gamma$ may be polarization-suppressed because the numerator of each terms is small. In this case, the wave vectors and polarization vectors are at orthogonal angles, so that vector inner products in $\Theta^s$ and $\Phi^s$ are small. Finally, $\Gamma$ may be energy-suppressed because its denominator is large. In this case, a large fraction of the wave energy is kinetic or thermal, so the electromagnetic field amplitude is small for given wave energy.

A consequence of the three-wave interactions is the parametric decay instability. During the instability, a large amplitude pump wave $a_1$ decays to the frequency-downshifted daughter waves $a_2$ and $a_3$, whose relative phases are automatically locked. In the linear regime of parametric interaction, $a_1$ barely changes and $a_2$ and $a_3$ grow almost exponentially with growth rate
\begin{equation}
    \label{eq:growth}
    \gamma_0=\frac{|\Gamma a_1|}{\sqrt{\omega_2\omega_3}}.
\end{equation}
The actual growth rate observed in experiments are likely influenced by wave damping, which includes both collisional and collisionless damping. Damping effects can be important and may be introduced phenomenologically in the three-wave equations. However, the ideal-fluid model does not captured damping self-consistently.

The growth rate may be compared to that of Raman backscattering $\gamma_R=\sqrt{\omega_1\omega_p}|a_1|/2$ in cold unmagnetized plasmas of the same density, where $\omega_p^2=\sum_s\omega_{ps}^2$ is the total plasma frequency. We can write the growth rate $\gamma_0=\gamma_R\mathcal{M}$, then
\begin{equation}
    \label{eq:growthM}
    \mathcal{M}=2\frac{|\Gamma|}{\omega_p^2}\Big(\frac{\omega_p^3}{\omega_1\omega_2\omega_3}\Big)^{1/2}.
\end{equation}
The normalized growth rate is now symmetric with respect to the three waves, and is proportional to the coupling coefficient up to some kinematic factors. Since $\Gamma\sim\omega_p^2$, the normalized growth rate is zero in the limit $\omega_p\rightarrow 0$. This is expected because there is no three-wave coupling in the vacuum.

\section{Examples\label{sec:example}}
\begin{figure}[b]
    \centering
    \includegraphics[width=0.48\textwidth]{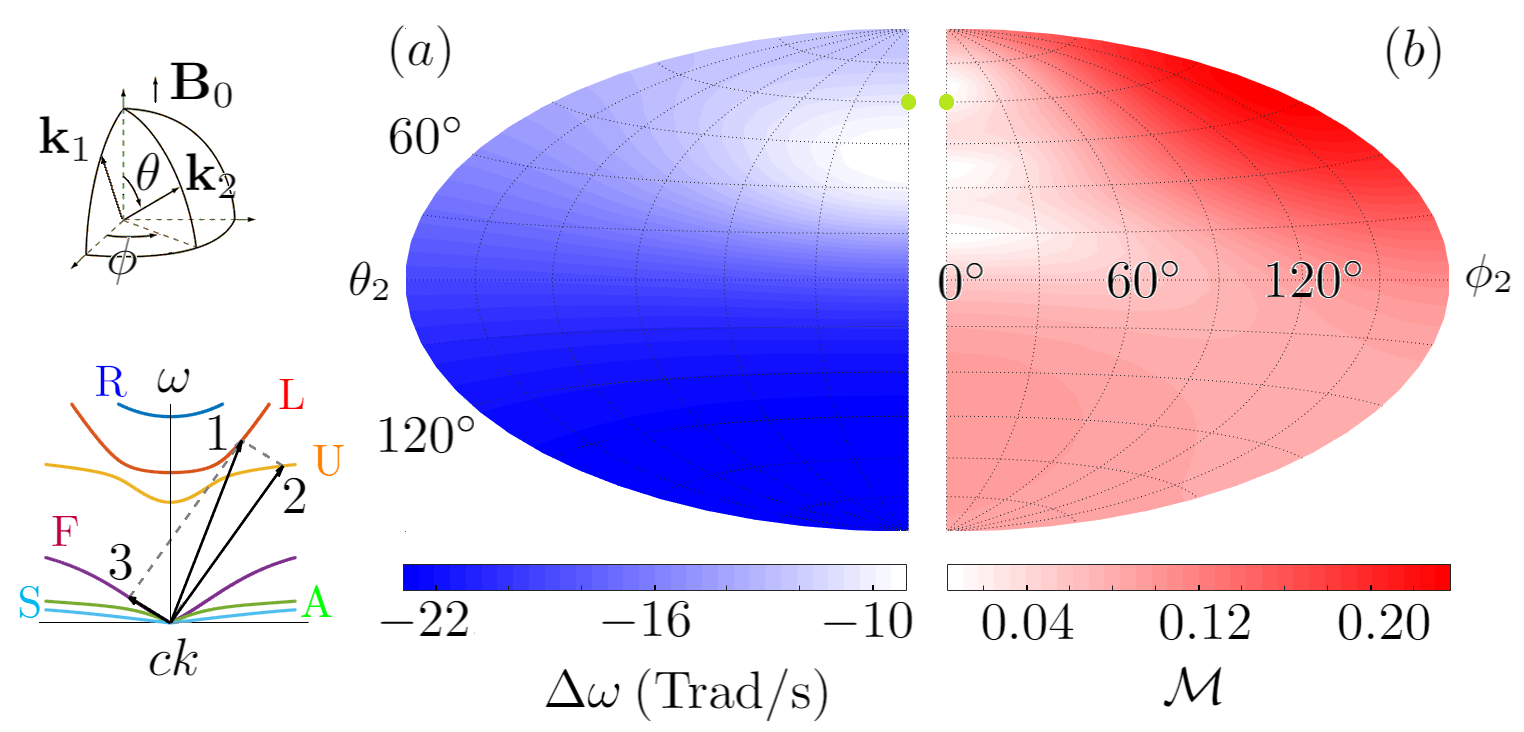}
    \caption{Maps of frequency downshift (a) and normalized growth rate (b) when an L wave decays to U and F daughter waves. Plasma parameters are the same as in Fig.~\ref{fig:wave}. The pump wave (green dot) has frequency \mbox{$\omega_1=75$ Trad/s}, and propagates at $\theta_1=30^\circ$ with respect to $\mathbf{B}_0$. The U-daughter wave propagates at polar angle $\theta_2$ and azimuthal angle $\phi_2$. Due to the presence of $\mathbf{B}_0$, backscattering is not the strongest. Moreover, special angles exist where the coupling is suppressed.}
    \label{fig:map}
\end{figure}

The above general theory is applicable to a discrete spectrum of weakly-damped and weakly-coupled waves in magnetized warm-fluid plasmas. The waves can propagate in any directions and have arbitrary frequencies.
An example of resonant interaction is shown in Fig.~\ref{fig:map}, where $a_1$ is on the L branch, $a_2$ is on the U branch, and $a_3$ is on the F branch. 
The matching of resonance conditions in the Fourier space is shown in the inset for collimated scattering. Also shown in the inset is the interaction geometry in the configuration space, where $\langle\hat{\mathbf{k}}_1,\mathbf{B}_0\rangle$ is fixed at $30^\circ$, while $\hat{\mathbf{k}}_2$ has polar angle $\theta_2$ and azimuthal angle $\phi_2$. 
Due to mirror symmetry, the frequency downshift $\Delta\omega=\omega_2-\omega_1$ is plotted only for the western hemisphere (Fig.~\ref{fig:map}a), while the normalized growth rate $\mathcal{M}$ is shown only for the eastern hemisphere (Fig.~\ref{fig:map}b).
Plasma parameters used in this example are the same as in Fig.~\ref{fig:wave}. The pump wave (marked by green dot) has frequency \mbox{$\omega_1=75$ Trad/s}, which corresponds to $ck_1\approx 51.54$ Trad/s. 
Due to the presence of the magnetic field, the coupling has intricate angular dependence.

A numerically robust procedure for evaluating the coupling coefficient and the growth rate is as follows. 
First, imagine we launch a pump wave with frequency $\omega_1$ in direction $\hat{\mathbf{k}}_1$ on a given branch. Then, the wave number $k_1$ can be solved from the dispersion relation using procedures described in Sec.~\ref{sec:dispersion}, and the unit polarization vector $\mathbf{e}_1$ can be computed using procedures described in Sec.~\ref{sec:polarization}. The matrix representation of the warm forcing operator $\hat{\mathbb{F}}_1$ can be computed, based on which the wave energy operator $\mathbb{H}_1$ and the wave energy coefficient $u_1$ can be evaluated.
Second, suppose we place a detector along $\hat{\mathbf{k}}_2$, we can in principle detect all waves whose wave number is such that $\omega_1-\omega_2(\mathbf{k}_2)=\omega_3(\mathbf{k}_1-\mathbf{k}_2)$ is in resonance with a third wave. For each pair of wave branches, solving the above resonance condition gives $k_2$, from which $\omega_2$ and $\mathbf{e}_2$ can be determined and $\hat{\mathbb{F}}_2$ and $u_2$ can be evaluated. At the same time, $\mathbf{k}_3$ also becomes known, from which $\omega_3$, $\mathbf{e}_3$, $\hat{\mathbb{F}}_3$ and $u_3$ can be evaluated.
Finally, once these quantities are determined, the scattering strengths can be evaluated, so are the coupling coefficient [Eq.~(\ref{eq:coupling})] and the growth rate [Eq.~(\ref{eq:growthM})]. 
This procedure is used to obtain Fig.~\ref{fig:map}, and will be further demonstrated below using two examples.

\subsection{Scattering of high-frequency lasers\label{sec:laser}}
When the pump $a_1$ and the seed $a_2$ are high-frequency EM waves, whose frequencies $\omega_1,\omega_2\gg \omega_{ps},|\Omega_s|$, they asymptote to vacuum EM waves. In the high-frequency limit, $\beta,\hat{\beta}\rightarrow 0$, $\gamma,\hat{\gamma}\rightarrow 1$, and the forcing operators $\hat{\mathbb{F}}_1, \hat{\mathbb{F}}_2\sim\mathbb{I}$. 
Since $a_3$ is a plasma wave with much lower frequency, the electromagnetic scattering is dominantly $\Theta^s\simeq -(c\mathbf{k}_3\cdot\mathbf{f}^*_{s,3})(\mathbf{e}_1\cdot\mathbf{e}_2^*)/\omega_3$
whenever $\mathbf{e}_1\cdot\mathbf{e}_2^*$ is of order unity. 
Moreover, since vacuum EM waves are transverse, the thermal scattering becomes $\Phi^s\simeq u_s^2 k_3^2 (c\mathbf{k}_3\cdot\mathbf{f}^*_{s,3})(\hat{\mathbf{k}}_3\cdot\mathbf{e}_1)(\hat{\mathbf{k}}_3\cdot\mathbf{e}_2^*)/\omega_1\omega_2\omega_3$.
When the angle between $\mathbf{k}_1$ and $\mathbf{k}_2$ is not too small, $ck_3\sim \omega_1, \omega_2$ is always large, so $\mathbf{e}_3\sim\hat{\mathbf{k}}_3$ is approximately longitudinal.
Then, $\mathbb{P}_3\mathbf{e}_3\simeq \hat{\gamma}^2_3\mathbf{e}_3$ and $\mathbf{e}_3\cdot\mathbf{f}_3^*\simeq \gamma^2_3 \hat{\gamma}^2_3(1-\beta_3^2\cos^2\theta_3)$, where $\theta_3=\langle\mathbf{k}_3,\mathbf{B}_0\rangle$ and the thermal ratio $\hat{\beta}_3^2\simeq u^2k_3^2\gamma_3^2(1-\beta_3^2\cos^2\theta_3)/\omega_3^2$.
Finally, the wave energy coefficients $u_1, u_2\simeq 1$, and $u_3$ can be evaluated with 
$\mathbf{f}_3\!\cdot\!\mathbf{f}_3^*=\hat{\gamma}_3^4[\cos^2\theta_3+\gamma_3^4(1+\beta_3^2)\sin^2\theta_3]$.
Using these asymptotics, the coupling coefficient and the parametric growth rate can be approximated.

The above approximations clearly recover the cold magnetized case \cite{Shi2017three}, and they also recover the well-known Raman and Brillouin scatterings in warm unmagnetized plasmas. In the unmagnetized limit, $\beta=0$, $\gamma=1$, and $\hat{\beta}_3^2= u^2k_3^2/\omega_3^2$. 
Moreover, the plasma waves are purely longitudinal with the dispersion relation $\omega_3^2=\sum_s \omega_{ps}^2 \hat{\gamma}_{s,3}^2$. Then, $\mathbf{e}_3\cdot\mathbf{f}_3^*= \hat{\gamma}^2_3$, $\mathbf{f}_3\!\cdot\!\mathbf{f}_3^*=\hat{\gamma}_3^4$, and $u_3=\sum_s\omega_{ps}^2\hat{\gamma}_{s,3}^4/\omega_3^2$. 
In most cases $\Phi^s\ll\Theta^s$, because $\omega_3\ll\omega_1,\omega_2\sim ck_3$.
For the same reason, $k_3\simeq 2k_1\sin(\alpha/2)$, where $\alpha$ is the angle between $\mathbf{k}_1$ and $\mathbf{k}_2$. 
Now let us focus on quasi-neutral electron-ion plasma with $Z_i=1$. In this two-species plasma, there are two longitudinal waves.
The high-frequency wave is the Langmuir wave, whose mediation gives rise to the Raman scattering. Since \mbox{$\lambda_D k_3\ll 1$} is required for weak collisionless damping, 
the dispersion relation is \mbox{$\omega_3^2\simeq\omega_p^2$}, so $\hat{\beta}_{i,3}^2,\hat{\beta}_{e,3}^2\ll1$, and $\hat{\gamma}_{i,3}^2\simeq\hat{\gamma}_{e,3}^2\simeq1$. The normalized unmagnetized Raman growth rate is then
\begin{equation}
    \label{eq:MRaman}
    \mathcal{M}_R\simeq \sin\frac{\alpha}{2} \Big(\frac{\omega_p}{\omega_3}\Big)^{1/2}\Big(1-\frac{1}{M_i}\Big)|\mathbf{e}_1\cdot\mathbf{e}_2^*|.
\end{equation}
We see responses by the two species destructively interfere, and exact cancellation occurs in electron-positron plasma where $M_i=1$.
On the contrary, the responses constructively interfere for the low-frequency sound wave, whose mediation gives rise to the Brillouin scattering. 
For sound wave, the dispersion relation is $\omega_3^2\simeq c_s^2k_3^2$, where the sound speed $c_s^2=2M_iu_i^2/(M_i+1)$ assuming $u_e^2=M_iu_i^2$. Then, $\hat{\gamma}_e^2\simeq2/(1-M_i)$ and $\hat{\gamma}_i^2\simeq2M_i/(M_i-1)$ are of opposite signs. The wave energy coefficient \mbox{$u\simeq 4\omega_p^2M_i/\omega^2(M_i-1)^2$},
and the unmagnetized Brillouin growth rate is then
\begin{equation}
    \label{eq:MBrillouin_ei}
    \mathcal{M}_B\simeq \sin\frac{\alpha}{2} \Big(\frac{\omega_p}{M_i\omega_3}\Big)^{1/2} \,|\mathbf{e}_1\cdot\mathbf{e}_2^*|,
\end{equation}
for both electron-ion and electron-positron plasmas when temperature is not too high.
The above recovers the weak-coupling results in the literature \cite{Sjolund1967parametric,Gorbunov1969perturbation,Litvak1971induced,Forslund1973nonlinear,Drake1974parametric,Forslund1975theory,Edwards2016strongly}, which were derived for unmagnetized plasmas in the parametric interaction picture.

Without any approximation, the growth rate 
can be evaluated numerically to determine collective laser scattering in magnetized plasmas. 
Consider an example relevant to laser-driven magnetized liner fusion \cite{Davies2017laser,Barnak2017laser}. In the experimental design, a preheat laser with 351-nm wavelength propagates along $\mathbf{B}_0$ of about 30 T. The $\text{D}_2$ plasma, for which $Z_i=1$ and $M_i\approx 3671$, has density $\sim$1.5 mg/$\text{cm}^3$. After fully ionized, the number density is about $n_e=n_i= 4.5\times 10^{20}\,\text{cm}^{-3}$, and the plasma temperature is about \mbox{$T_e=400$ eV} and \mbox{$T_i=150$ eV}. 
In this example, the Debye length \mbox{$\sim 10^{-2}\,\mu$m} is much smaller than the laser wavelength $\lambda_0$, and $\lambda_0$ is much smaller than the collisional mean free path \mbox{$\sim 10\,\mu$m}, so the ideal fluid model is applicable. Moreover, since $\lambda_0$ is much smaller than the ion gyro radius, ions are essentially unmagnetized and the A branch has minuscule contribution. Therefore, effects of magnetization are mainly due to electrons.

\begin{figure}[t]
    \centering
    \includegraphics[width=0.48\textwidth]{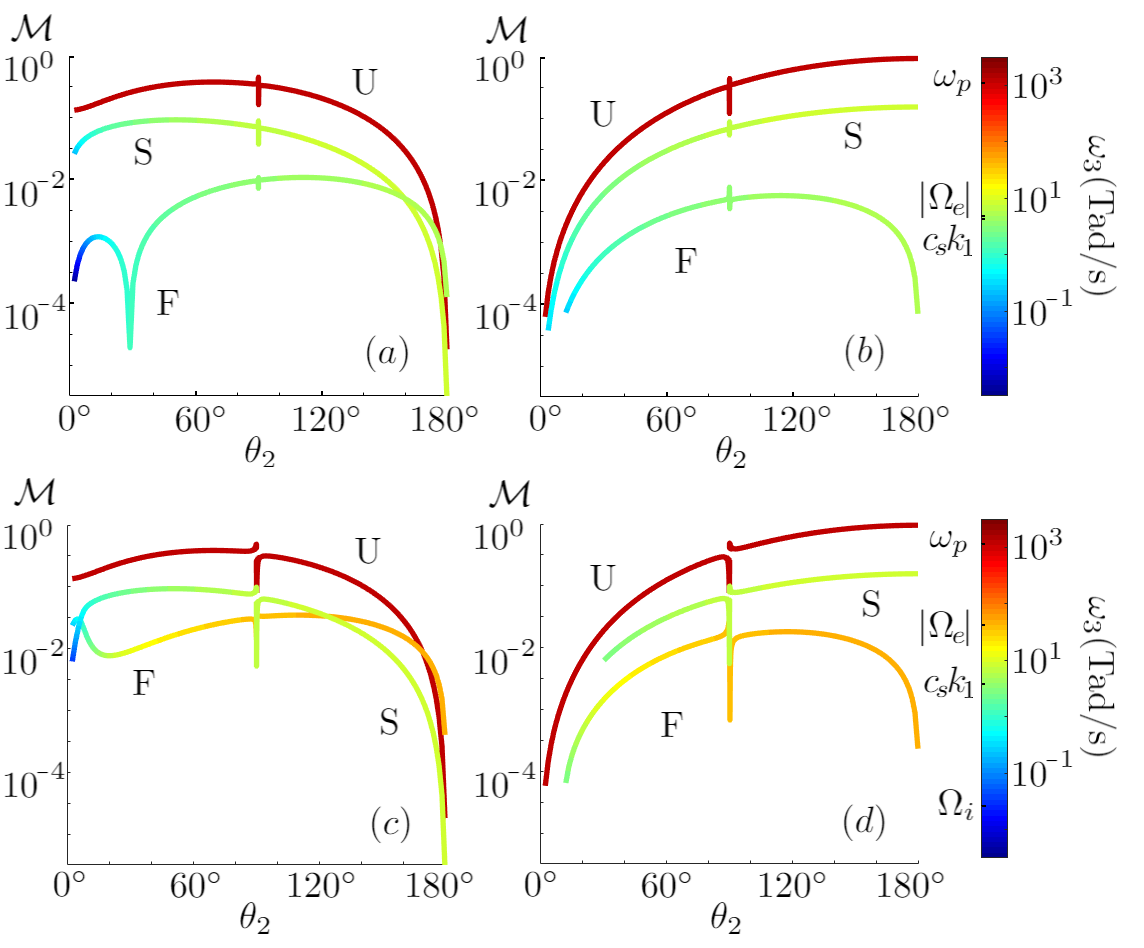}
    \caption{Decay rates of a pump laser in a deuterium plasma via R$\rightarrow$R (a, c) and R$\rightarrow$L (b, d) scattering . The rates are in units of Raman backscattering, and the curves are color-coded by frequency downshifts. The 351-nm pump laser propagates along a 30-T magnetic field (a, b) or a 300-T field (c, d), and the scattered light propagates at angle $\theta_2=\langle \mathbf{k}_2, \mathbf{B}_0\rangle$. The plasma density is $n_e=n_i= 4.5\times 10^{20}\,\text{cm}^{-3}$, the temperature is \mbox{$T_e=400$ eV} and \mbox{$T_i=150$ eV}, and the polytropic index $\xi_e=\xi_i=3$. Since $|\Omega_e|\ll\omega_p$, scattering mediated by the U branch is close to Raman; since $v_A<c_s$, scattering mediated by the S branch is close to Brillouin. Additionally, the laser can scatter from the F branch, which is energy-suppressed in weak magnetic field.}
    \label{fig:laser}
\end{figure}

Due to cylindrical symmetry, the scattering only depends on the polar angle $\theta_2$, which is $0^\circ$ for forward scattering and $180^\circ$ for backward scattering. 
The growth rates, in units of Raman backscattering, are shown in Fig.~\ref{fig:laser}, where the curves are color-coded by frequency downshifts.
The growth rates are polarization-dependent, and the eigenmodes are elliptically polarized, except when $\theta=90^\circ$ where they become the linearly polarized X and O wave. For the R-wave pump, scattering to the R branch (a, c) is polarization-suppressed for near backward scattering, while scattering to the L branch (b, d) is polarization-suppressed for near forward scattering.
When \mbox{$B_0=30$ T} (a, b), \mbox{$\Omega_e\approx 5.3$ Trad/s} is much smaller than \mbox{$\omega_p\approx 1.2\times 10^3$ Trad/s}, so scattering from the U branch is close to Raman. 
Similarly, since $v_A/c\approx 7\times 10^{-5}$ is smaller than $c_s/c\approx 9\times 10^{-4}$, the sound wave is little modified, and scattering from the S branch is close to Brillouin. 
Other than modifying Raman and Brillouin, the magnetic field introduces additional modes from which the laser can scatter. However, in weak magnetic fields, scattering from the F branch is energy-suppressed, because the F branch is dominated by electron cyclotron motion. 
In larger magnetic fields, for example \mbox{$B_0=300$ T} (c, d), effects of magnetization then become larger.

\subsection{Scattering of MHD waves\label{sec:MHD}}
To illustrate that the general formula is applicable to any wave triplets, let us consider scattering of MHD waves as another example. In this case, the wave frequency $\omega\ll\Omega_i$, and the fluid-Maxwell's equations asymptote to MHD equations. Consequently, the wave dispersion relation also asymptotes to that of the MHD waves (Appendix~\ref{app:low}). The asymptotics are particularly simple for wave propagation parallel to $\mathbf{B}_0$, where the forcing operator becomes $\hat{\mathbb{F}}_\parallel\mathbf{Z}=\gamma^2(\mathbf{Z}+i\beta\mathbf{Z}\times\mathbf{b})+(\hat{\gamma}^2-\gamma^2)(\mathbf{Z}\cdot\mathbf{b})\mathbf{b}$. It may be tempting to already take the $\omega/\Omega\rightarrow 0$ limit for $\hat{\mathbb{F}}_\parallel$. However, the limit should be taken only after summations over species is carried out, because leading terms may turn out to cancel. 

First, let us determine the approximate wave dispersion relations. 
Consider two species plasmas with $Z_i=1$, then the sum in $\mathbb{D}_{11}$ is $\omega_{pe}^2\gamma_e^2+\omega_{pi}^2\gamma_i^2\simeq -c^2\omega^2/v_A^2$, where $v_A^2=c^2M_i\Omega_i^2/\omega_p^2$ is the Alfv\'en speed. 
The sum in $\mathbb{D}_{12}$ is $\beta_e\omega_{pe}^2\gamma_e^2+\beta_i\omega_{pi}^2\gamma_i^2\simeq -(1-1/M_i)c^2\omega^3/v_A^2\Omega_i$.
Finally, the sum in $\mathbb{D}_{33}$ is $\omega_{pe}^2\hat{\gamma}_e^2+\omega_{pi}^2\hat{\gamma}_i^2\simeq \omega^2\omega_p^2(\omega^2-c_s^2k^2)/(\omega^2-u_e^2k^2)(\omega^2-u_i^2k^2)$. 
The dispersion tensor $\mathbb{D}$ for parallel wave propagation can then be easily determined in the field coordinate. The longitudinal wave satisfies $\omega^2\simeq c_s^2k^2$, 
which is essentially the unmagnetized sound wave. 
The transverse waves satisfy $(1+v_A^2/c^2)\omega^2=v_A^2k^2\pm(1-1/M_i)\omega^3/\Omega_i$. The ``+" branch has higher phase velocity and is right-handed circularly polarized with $\mathbf{e}\propto(1,i,0)$; the ``--" branch has lower phase velocity and is left-handed circularly polarized with $\mathbf{e}\propto(1,-i,0)$. To lowest order in $\omega/\Omega_i$, the two branches merge into dispersionless Alfv\'en waves $\omega^2=c_A^2k^2$, 
where $c_A^2=v_A^2/(1+v_A^2/c^2)$. The energy coefficient of the Alfv\'en waves is $u\simeq c^2/c_A^2$, which is usually very large because most wave energy is contained in magnetic and fluid fluctuations instead of the wave electric field. In the MHD limit $v_A^2 \ll c^2$, the above results recover the dispersion relations of parallel-propagating MHD waves.

Now we can compute three-wave coupling between MHD waves. For parallel wave propagation, other than the coupling between three sound waves, which is discussed in Appendix \ref{app:neutral},  the only nonzero coupling is between two Alfv\'en waves of the same polarization ($a_1, a_2$) and a sound wave ($a_3$).
Since the waves are dispersionless, the resonance conditions can be satisfied only when $a_1$ and $a_2$ are counter propagating. In this geometry, the resonant wave vectors are $k_2/k_1=|c_s-c_A|/(c_s+c_A)$ and $k_3/k_1=2c_A/(c_s+c_A)$. 
To compute the the scattering between these waves, notice that for the sound wave $\mathbf{f}=\hat{\gamma}^2\mathbf{b}$, while for the Alfv\'en wave $\mathbf{f}=\mathbf{e}/(1\pm\beta)$ where $\pm$ corresponds to the L and R polarizations.
The electromagnetic scattering due to each species is thereof $\Theta_s\simeq -[1/(\omega_1+\Omega_s)+1/(\omega_2+\Omega_s)]c\omega_1\omega_2\omega_3/c_A(\omega_3^2-u_s^2k_3^2)$, and the thermal scattering is $\Phi_s\simeq 0$ because two waves are transverse. Summing over species and then take the limit $\omega/\Omega_i\rightarrow 0$, the coupling coefficient is
\begin{equation}
    \label{eq:couplingMHD}
    \Gamma_\parallel \simeq \frac{c_A^2}{v_A c_s} \frac{\omega_1\omega_2\omega_3}{4M_i\Omega_i}.
\end{equation}
The coupling 
can also be expressed in terms of the magnetic field
$a_1=M_i\Omega_i \mathcal{B}_1/\omega_1 B_0$ whereby the growth rate [Eq.~(\ref{eq:growth})] can be readily evaluated, which agrees with the weak-coupling result in the literature \cite{Jayanti1993parametric,Matsukiyo2003parametric,Wong1986parametric,Derby1978modulational,Goldstein1978instability} for both electron-ion and electron-positron plasmas.

\begin{figure}[t]
    \centering
    \includegraphics[width=0.48\textwidth]{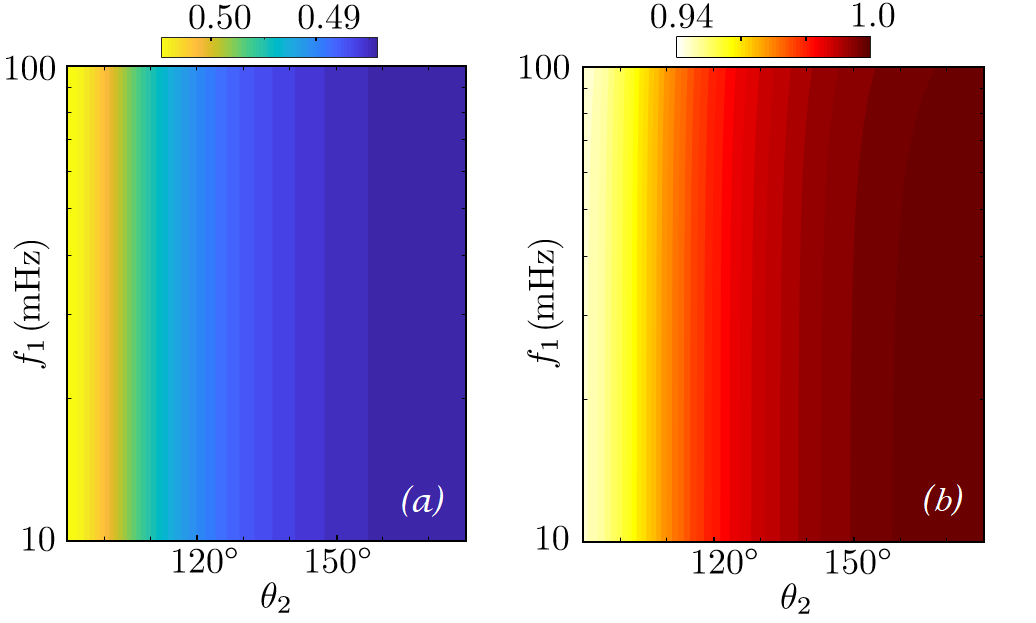}
    \caption{Resonant coupling between two Alfv\'en waves via the sound wave in solar corona type plasma with $c_s/v_A\approx0.35$. The pump Alfv\'en wave propagates along $\mathbf{B}_0$ with frequency $f_1$, while the daughter Alfv\'en wave propagates obliquely at angle $\theta_2$. The frequency of the daughter wave $f_2/f_1$ (a) and the coupling coefficient $\Gamma/\Gamma_\parallel$ (b) have weak dependence on both $\theta_2$ and $f_1$. Consequently, the parametric decay rate is only slightly larger for exact backward scattering.}
    \label{fig:AAS}
\end{figure}

Without any approximation, the exact formula of the resonant coupling coefficient can be evaluated numerically. Let us consider an example relevant for solar corona at a height comparable to the solar radius \cite{Phillips1995ulysses,Guhathakurta1996large,Aschwanden2006physics,Tomczyk2007alfven}. 
There, the plasma is mostly hydrogen with $M_i\approx 1837$. The plasma density \mbox{$n_e\sim n_i\sim 10^7\,\text{cm}^{-3}$}, the plasma temperature \mbox{$T_e\sim T_i\sim 100$ eV}, and the magnetic field \mbox{$B_0\sim$ 1 G}. Correspondingly, \mbox{$\omega_p\approx1.8\times10^8$ rad/s}, $c_s/c\approx 8.0\times 10^{-4}$, and $v_A/c\approx 2.3 \times 10^{-3}$. 
The ion cyclotron frequency \mbox{$\Omega_i\approx 10^4$ rad/s} is much higher than the observed Alfv\'en wave frequency, which is in the mHz band. In this frequency range, the low-frequency waves are well-approximated by ideal MHD waves.
Consider the coupling between two Alfv\'en waves via the sound wave.
The daughter wave frequency $f_2/f_1$ is shown in Fig.~\ref{fig:AAS}a, and the coupling coefficient $\Gamma/\Gamma_\parallel$ is shown in Fig.~\ref{fig:AAS}b, where $\Gamma_\parallel$ is given by Eq.~(\ref{eq:couplingMHD}).
While the coupling has strong dependence on $c_s/v_A$, it has very weak dependence on the frequency of the parallel pump Alfv\'en wave. Moreover, the dependence on $\theta_2$, the angle of the daughter wave with respect to the local magnetic field, is also weak.
Consequently, the decay of the parallel pump wave only slight prefers exact backward geometry.

\section{Discussion and summary \label{sec:discussion}}
Beyond linear waves, this paper treats coherent three-wave interactions in magnetized plasma by solving the warm-fluid model to second order. Unlike previous attempts, which were specialized for each wave triad in restricted geometry, here, the systematic treatment using perturbation theory offers a unified description of all possible interactions at arbitrary angles. This methodology, first introduced for magnetized cold-fluid plasma \cite{Shi2017three}, is extended to incorporate thermal effects. The ideal-fluid model is applicable when all wavelengths are much larger than the Debye length, while much shorter than the collisional mean free path. In this regime, thermal effects enter indirectly through the forcing operator [Eq.~(\ref{eq:FhatZ})], as well as directly in the quadratic response [Eq.~(\ref{eq:Rpq})]. Nevertheless, the second-order electric-field equation [Eq.~(\ref{eq:E2eq})] remains formally unchanged. 

The formalism developed in this paper is not only general, but also practical, whereby numerical values of the coupling coefficient can be obtained. The coupling coefficient is an essential parameter in the commonly-used three-wave equations [Eqs.~(\ref{eq:da1})-(\ref{eq:da3})]. Previously, little is known about the numerical value of the coupling coefficient when the plasma becomes magnetized. Now, a general formula has been provided [Eq.~(\ref{eq:coupling})], which can be evaluated for any three resonantly interacting waves.
To demonstrate the powerfulness of the general formula, the coupling between high-frequency lasers via Raman [Eq.~(\ref{eq:MRaman})] and Brillouin [Eq.~(\ref{eq:MBrillouin_ei})] scatterings in unmagnetized plasmas are recovered as special cases. 
Moreover, the same formula also recovers coupling between two Alfv\'en waves and a sound wave [Eq.~(\ref{eq:couplingMHD})], which are at the other extreme of the wave spectrum.
While asymptotic expressions of the general formula may be found for special cases, the exact formula can always be evaluated using numerical procedures demonstrated in this paper. 
Based on nontrivial analytic simplifications, numerical evaluations the of coupling coefficient can now be made efficient and robust.

In summary, this paper derives a general formula governing resonant three-wave interactions in magnetized warm-fluid plasmas in the weak-coupling regime. 
Applying the formula to magnetized inertial confinement fusion conditions, the magnetic field is found to modify Raman and Brillouin scatterings of lasers, as well as introduce additional scattering modes at oblique angles.
For parameters relevant to solar corona, the formula for parallel coupling between two Alfv\'en waves via the sound wave is found to give good approximations also at oblique angles. Due to weak angular dependence, exact backscattering is only slightly preferred over oblique decays.

\begin{acknowledgments}
The author thanks John D. Moody, Bradley B. Pollock, David J. Strozzi, Pierre A. Michel, Nathaniel J. Fisch, Matthew R. Edwards, and Matthew W. Kunz for helpful discussions. 	
This work was performed under the auspices of the U.S. Department of Energy by Lawrence Livermore National Laboratory under Contract DE-AC52-07NA27344 and was supported by the Lawrence Fellowship through LLNL-LDRD Program under Project No. 19-ERD-038.
\end{acknowledgments}

\appendix
\section{Asymptotic dispersion $ck\rightarrow 0$\label{app:low}}
The dispersion relation contains gapped and gapless modes. For gapped modes, the wave frequency $\omega\rightarrow \omega_c$ when $ck\rightarrow 0$, where $\omega_c$ is some finite cutoff frequency. For gapless modes, $\omega\rightarrow 0$ when $ck\rightarrow 0$, but the refractive index $n=ck/\omega$ approaches some finite constant. 
The asymptotic dispersion relation is useful for analytic approximations, and may be used as initial guesses for numerical root finding.

For gapped modes, the cutoff frequencies are solutions of $C(\omega_c)=0$, where $C$ is given by Eq.~(\ref{eq:C}). Since thermal effects vanish, one cutoff frequency is always $\omega_p$. The other cutoff frequencies are solutions of $R(\omega_c)=L(-\omega_c)=0$. In a magnetized plasma of $N_s$ species, there are $N_s+1$ non-negative solutions, which becomes strictly positive when the plasma is not quasi-neutral. 
For finite but small $ck$, we can expand near $\omega_c$. The asymptotic dispersion relation is quadratic: $\omega\simeq\omega_c^2+\delta\omega^2$, where $\delta\omega^2=2 B c^2k^2/\omega_c\partial_\omega C$.
Here, $B$ and $\partial_\omega C$ are evaluated at $\omega=\omega_c$ and $ck=0$. The analytic expression is simple, since thermal effects vanish.

On the other hand, thermal effects are important for gapless modes. To obtain asymptotic dispersion relation when $\omega\rightarrow 0$, we can expand using Laurent series.
After tedius but otherwise straightforward expansions, the leading terms in a quasi-neutral plasma are
\begin{eqnarray}
    \label{eq:w2A}
    \omega^2 A&\simeq& -I_2 c^2_\theta,\\
    \label{eq:w2B}
    \omega^2 B&\simeq& (I_0I_2-I_1^2) s^2_\theta-\frac{2 c^2}{c_A^2}I_2,\\
    \label{eq:w2C}
    \omega^2 C&\simeq& \Big[(I_0I_2-I_1^2) s^2_\theta-\frac{c^2}{c_A^2}I_2\Big]\frac{c^2}{c_A^2c^2_\theta},
\end{eqnarray}
assuming $c_\theta^2\gg \omega^2/\Omega^2$.
The dispersion coefficients
$I_0=1+\sum_s \omega_{ps}^2\eta_s^2/\Omega_s^2$, $I_1=\sum_s \omega_{ps}^2\eta_s^2/\Omega_s$, and $I_2=\sum_s \omega_{ps}^2\eta_s^2$, where $\eta_s^2=1/(1-n^2c_\theta^2u_s^2/c^2)$.
In the cold limit, $I_0\rightarrow c^2/c_A^2$, $I_1\rightarrow 0$, and $I_2\rightarrow \omega_p^2$.
Substituting Eqs.~(\ref{eq:w2A})-(\ref{eq:w2C}) into the dispersion relation [Eq.~(\ref{eq:disp})], we obtain an equation for $n^2$. The Alfv\'en wave decouples with the dispersion relation
\begin{equation}
    \omega^2=c_A^2 k^2\cos^2\theta.
\end{equation}
What remains are the fast wave mixed with the sound waves, which is given by 
\begin{equation}
    \label{eq:FastSounds}
    I_2\Big(n^2\cos^2\theta-\frac{c^2}{c_A^2}+I_0\sin^2\theta\Big)=I_1^2\sin^2\theta.
\end{equation}
A special case is when all species are cold. Then, the sound wave vanishes and the above recovers the cold fast wave $\omega^2=c_A^2 k^2$. 
In more general cases, a numerically robust procedure for solving the dispersion relation is to remove poles of Eq.~(\ref{eq:FastSounds}), and convert it to a polynomial of $n^2c_\theta^2$ of degree $N_c$+1. 
When $N_t\ge N_s-1$, the leading coefficient is $\sum_s\omega_{ps}^2\prod_{s'\ne s}(-\mu_{s'}^2)$, otherwise the leading coefficient is $\prod_{\mu_{s'}\ne 0}(-\mu_{s'}^2)\cdot\sum_{\mu_{s'}=0}\omega_{ps}^2$.
The polynomial equation has exactly $N_c$+1 real and positive roots, which can be found by standard numerical methods.
The above is a multi-fluid extension of MHD, which retains only one sound wave.  

\section{Three-wave in neutral fluid\label{app:neutral}}
To illustrate that that turbulent and thermal beatings are originated from fluid nonlinearities, let us consider three-wave interactions in neutral fluid, which is described by
\begin{eqnarray}
    \label{eq:continuity_fluid}
    &&\partial_t\rho+\nabla\cdot(\rho \mathbf{v})=0,\\
    &&\rho d_t \mathbf{v}=-\nabla p,\\
    &&\rho d_t p=\xi p d_t\rho,
\end{eqnarray}
where $\rho$ is the mass density and $d_t=\partial_t+\mathbf{v}\cdot\nabla$ is the convective derivative at the fluid velocity. 

The linearized fluid equations describe sound waves. Suppose we weakly perturbe the equilibrium with constant $\rho_0$, $p_0$, and $\mathbf{v}_0=\mathbf{0}$, the first-order fluid velocity is $\mathbf{v}_1=\frac{1}{2}\sum_{\mathbf{k}\in\mathbb{K}_1}\exp(i\theta_\mathbf{k})\mathbfcal{V}_{1,\mathbf{k}}$.
The continuity equation then gives $\rho_1/\rho_0=\frac{1}{2}\sum_{\mathbf{k}\in\mathbb{K}_1} \exp(i\theta_\mathbf{k})\mathbf{k}\cdot\mathbfcal{V}_{1,\mathbf{k}}/\omega_\mathbf{k}$,
and the pressure equation gives $p_1=u^2\rho_1$, where $u^2=\xi p_0/\rho_0$ is the thermal speed. 
Substituting these into the momentum equation, each Fourier amplitude satisfies
\begin{equation}
   \label{eq:v1_fluid}
    (\omega_\mathbf{k}^2-u^2\mathbf{k}\mathbf{k})\mathbfcal{V}_{1,\mathbf{k}}=\mathbf{0}.
\end{equation}
The dispersion operator is now
$\bar{\mathbb{D}}_\mathbf{k}=\omega_\mathbf{k}^2-u^2\mathbf{k}\mathbf{k}$. 
The eigenmode satisfies the dispersion relation $\omega^2=u^2\mathbf{k}^2$, and
is the longitudinally polarized sound wave.

To second order in multiscale perturbative analysis, the equations can be obtained from Eqs.~(\ref{eq:n2})-(\ref{eq:p2}) by setting the electromagnetic contributions to zero. The pressure equation 
gives $p_2$ by Eq.~(\ref{eq:p2wave}) after replacements $m n\rightarrow\rho$ and $\varepsilon/m\rightarrow u^2$. Expanding the second-order velocity as $\mathbf{v}_{2}=\sum_{\mathbf{k}}\exp(i\theta_\mathbf{k})\mathbfcal{V}_{2,\mathbf{k}}/2$, the 
continuity equation gives $\rho_2$ by
Eq.~(\ref{eq:n2wave}) after replacing $ie\hat{\mathbb{F}}\mathbfcal{E}/m\omega\rightarrow\mathbfcal{V}$. 
Using $\bar{\mathbb{D}}_\mathbf{k}\mathbfcal{V}_{1,\mathbf{k}}=\mathbf{0}$, the second-order momentum equation can then be written as
\begin{eqnarray}
    \label{eq:v2eq_fluid}
    \nonumber
    &&\sum_{\mathbf{k}\in\mathbb{K}_2}\bar{\mathbb{D}}_\mathbf{k} \mathbfcal{V}_{2,\mathbf{k}} e^{i\theta_\mathbf{k}}\\
    &+&i\sum_{\mathbf{k}\in\mathbb{K}_1}
    \Big(\frac{\partial\bar{\mathbb{D}}_\mathbf{k}}{\partial\omega_\mathbf{k}}\partial_{t_1} 
    -\frac{\partial \bar{\mathbb{D}}_\mathbf{k}}{\partial\mathbf{k}}\cdot\nabla_1\Big)\mathbfcal{V}_{1,\mathbf{k}}e^{i\theta_\mathbf{k}} \\
    \nonumber
    &=&\frac{1}{2}\sum_{\mathbf{p},\mathbf{q}\in\mathbb{K}_1}(\omega_\mathbf{p}+\omega_\mathbf{q})\bar{\mathbf{S}}_{\mathbf{p},\mathbf{q}} e^{i\theta_\mathbf{p}+i\theta_\mathbf{q}} ,
\end{eqnarray}
which is formally identical to Eq.~(\ref{eq:E2eq}) if the later is written in terms of velocity perturbations. 
Analogously,
$\bar{\mathbf{S}}_{\mathbf{p},\mathbf{q}}=(\bar{\mathbf{R}}_{\mathbf{p},\mathbf{q}}+\bar{\mathbf{R}}_{\mathbf{q},\mathbf{p}})/2$, and the only difference is that now $\bar{\mathbf{R}}_{\mathbf{p},\mathbf{q}}=\bar{\mathbf{T}}_{\mathbf{p},\mathbf{q}}+\bar{\mathbf{U}}_{\mathbf{p},\mathbf{q}}$. Here, $\bar{\mathbf{T}}_{\mathbf{p},\mathbf{q}}$ and $\bar{\mathbf{U}}_{\mathbf{p},\mathbf{q}}$ can be obtained from Eqs.~(\ref{eq:Tbeat}) and (\ref{eq:Ubeat}) by replacing $\hat{\mathbb{F}}\mathbfcal{E}/\omega\rightarrow\mathbfcal{V}$. 
We see that turbulent and thermal beatings are intrinsically fluid nonlinearities.

The second-order velocity equation can be split into off-shell and on-shell equations. The off-shell equations can be solved by inverting the nondegenerate $\bar{\mathbb{D}}$, and the on-shell equations can be simplified using eigen projections. Suppose the resonance conditions are of the form ``$p=q+l$", the three-wave amplitude equations are 
\begin{eqnarray}
    d_t v_p=-\frac{i(1+\xi)}{4}k_p v_q v_l,
\end{eqnarray}
where $v_l=v_{-l}^*$ is the complex amplitude such that $\mathbfcal{V}_{1,\mathbf{k}_l}=v_l \hat{\mathbf{k}}_l$. 
In the parametric decay picture, the growth rate $\gamma_0=(1+\xi)(k_2k_3)^{1/2}|v_1|/4$. 
To obtain the above three-wave equations, I have used the fact that resonance conditions can be satisfied only when $\hat{\mathbf{k}}_1=\hat{\mathbf{k}}_2=\hat{\mathbf{k}}_3$, because the sound waves are dispersionless. 
Due to the special dispersion relation, three-wave interactions are one dimensional, along which any two copropagating waves can resonantly interact. 

\section{Eigen projection\label{app:onShell}}
To illustrate how the compatibility condition [Eq.~(\ref{eq:polCondition})] can be used in conjunction with the on-shell equation [Eq.~(\ref{eq:onShell})], let us consider unmagnetized cold plasma as an example. In this case, $\hat{\mathbb{F}}=\mathbb{I}$ is the identity operator and the dispersion tensor is
\begin{equation}
    \label{eq:D_unmag}
    \mathbb{D}=(\omega^2-\omega_p^2-c^2k^2)\mathbb{I}+c^2\mathbf{k}\mathbf{k}.
\end{equation}
The partial derivatives are $\partial\mathbb{D}/\partial\omega=2\omega\mathbb{I}$ and $\partial\mathbb{D}^{ij}/\partial k_l=c^2(k_i\delta_{jl}+k_j\delta_{il}-2k_l\delta_{ij})$. 
The on-shell equation is then of the form
\begin{equation}
    \label{eq:onShell_unmag}
    2\omega\partial_t\mathbfcal{E}=c^2[\mathbf{k}(\nabla\cdot\mathbfcal{E})+\nabla(\mathbf{k}\cdot\mathbfcal{E})-2(\mathbf{k}\cdot\nabla)\mathbfcal{E}]+\mathbf{S},
\end{equation}
where I have omitted the subscripts of $t_1$ and $\mathbf{x}_1$. Notice that the Eq.~(\ref{eq:onShell_unmag}) has redundant degrees of freedom, because the spatial derivatives originate from the projection operator $\mathbb{I}-\hat{\mathbf{k}}\hat{\mathbf{k}}$, which has a nontrivial kernel. 

For electromagnetic waves, the dispersion relation is $\omega^2=\omega_p^2+c^2k^2$. The dispersion tensor then becomes $\mathbb{D}=c^2\mathbf{k}\mathbf{k}$, which is a rank-1 operator. The null space is two dimensional, and the eigenmodes are transverse, which satisfy $\mathbf{k}\cdot\mathbfcal{E}=0$. The compatibility condition is satisfied if and only if
\begin{equation}
    \label{eq:polEM}
    c^2k^2\nabla\cdot\mathbfcal{E}+\mathbf{k}\cdot\mathbf{S}=0.
\end{equation}
Substituting the solution of $\nabla\cdot\mathbfcal{E}$ into Eq.~(\ref{eq:onShell_unmag}), the on-shell equation becomes
\begin{eqnarray}
    \label{eq:onShell_unmag_EM}
    (\partial_t+\frac{c^2\mathbf{k}}{\omega}\cdot\nabla)\mathbfcal{E} =\frac{\mathbf{S}^\perp}{2\omega},
\end{eqnarray}
where $c^2\mathbf{k}/\omega$ is nothing other than the group velocity,
and $\mathbf{S}^\perp=(\mathbb{I}-\hat{\mathbf{k}}\hat{\mathbf{k}})\mathbf{S}$ is the transverse projection. While the transverse projection is typically put in ``by-hand" when studying unmagnetized three-wave interactions \cite{Michel2014dynamic}, here I have shown why the projection necessarily arises. 

For the cold Langmuir waves, $\omega^2=\omega_p^2$, and $\mathbb{D}=c^2(\mathbf{k}\mathbf{k}-k^2\mathbb{I})$ is a rank-2 operator. The null space is thereof one dimensional, and the eigenmode is longitudinal, which satisfies $\mathbfcal{E}\parallelsum\mathbf{k}$. The compatibility condition is satisfied if and only if
\begin{equation}
    \label{eq:polLangmuir}
    (c^2\nabla\mathbf{k}\cdot\mathbfcal{E}+\mathbf{S})^\perp=\mathbf{0}.
\end{equation}
Substituting this into Eq.~(\ref{eq:onShell_unmag}), which can be separated into parallel and perpendicular components, the on-shell equation becomes
\begin{eqnarray}
    \label{eq:onShell_unmag_Langmuir}
    \partial_t\mathbfcal{E} =\frac{\mathbf{S}^\parallel}{2\omega},
\end{eqnarray}
As expected, the group velocity of the cold Langmuir wave is zero, and only the longitudinal component $\mathbf{S}^\parallel=\hat{\mathbf{k}}(\hat{\mathbf{k}}\cdot\mathbf{S})$ affects the wave evolution.

\bibliographystyle{apsrev4-1}

%

\end{document}